\documentclass[journal,draftclsnofoot,onecolumn,12pt,twoside]{IEEEtran}
\usepackage[T1]{fontenc}

\usepackage{amsmath, amsthm, amsfonts, amssymb, amsbsy}
\usepackage{setspace}
\usepackage{graphicx}
\usepackage{pstricks,arydshln}
\usepackage{algpseudocode}
\usepackage{algorithm}
\usepackage{multirow}
\usepackage{subfigure}
\usepackage{verbatim}
\usepackage{epstopdf}
\usepackage[noadjust]{cite}
\def\beginofproof{\noindent{\it Proof: }}
\newtheorem{theorem}{Theorem}
\newtheorem{lemma}{Lemma}

\newcommand\xqed[1]{%
  \leavevmode\unskip\penalty9999 \hbox{}\nobreak\hfill
  \quad\hbox{#1}}
\newcommand\fexmpl{\xqed{$\triangle$}}

\def\beginofproof{\noindent {\bf Proof. }}
\def\endofproof{\hfill$\Box$}

\newcommand{\Cref}[1]{Co\-ro\-lla\-ry\,\ref{#1}}

\newcommand{\cC}{{\cal C}}
\newcommand{\cP}{{\cal P}}

\begin{document}

\title{ Circular Buffer Rate-Matched Polar Codes}
\renewcommand{\markboth}[2]
{\renewcommand{\leftmark}{#1}\renewcommand{\rightmark}{#2}}

\author{
Mostafa El-Khamy, ~\IEEEmembership{Senior Member,~IEEE,}
Hsien-Ping Lin, ~\IEEEmembership{Student Member,~IEEE,}
        Jungwon Lee, ~\IEEEmembership{Senior Member,~IEEE, and Inyup Kang}
\thanks{This work was presented in part at the 2015 IEEE Global Communications Conference (GLOBECOM 2015), San Diego, CA, USA. }	}

\renewcommand\figurename{Fig.}
\maketitle
\begin{abstract}
A practical rate-matching system for constructing rate-compatible polar codes is proposed. The  proposed polar code circular buffer rate-matching is   suitable for transmissions on communication channels that support hybrid automatic repeat request (HARQ) communications, as well as for flexible resource-element rate-matching on single transmission channels. Our proposed circular buffer rate matching scheme  also incorporates a bit-mapping scheme for transmission on bit-interleaved coded modulation (BICM) channels using higher order modulations. An interleaver is derived from a puncturing order obtained with a low complexity progressive puncturing search algorithm on a base code of short length, and has the flexibility to achieve any desired rate at the desired code length, through puncturing or repetition. The rate-matching scheme is implied by a two-stage polarization, for transmission at any desired code length, code rate, and modulation order, and is shown to achieve the symmetric capacity of BICM channels. Numerical results on AWGN and fast fading channels show that the rate-matched polar codes have a competitive performance when compared to the spatially-coupled quasi-cyclic LDPC codes or LTE turbo codes, while having similar rate-dematching storage and computational complexities. 
\end{abstract}

\section{Introduction}
Polar codes and LDPC codes are under consideration for adoption in next generation wireless systems, such as the 3GPP new radio (NR) access technology \cite{3GPPNR}. One key component to enable adoption in next generation wireless systems is their flexible support of hybrid automatic repeat request (HARQ).  To design hybrid automatic repeat request (HARQ) rate-compatible (RC) codes for wireless systems, it is required that the information content and block length are the same across all codes with different rates. It is also required that the transmitted bits can be flexibly chosen according to the desired code rate and HARQ scheme,  such as Chase combining (CC) or incremental redundancy (IR). There have been significant research on designing HARQ RC LDPC codes, \cite{el2009design, DRCLDPC, RPLDC, SCQCLDPC}. It is still a challenging problem to design good flexible HARQ RC LDPC codes.

Polar codes have received much attention since they are the first class of error-correcting codes that provably achieve the binary symmetric capacity of memoryless channels with low-complexity encoders and decoders \cite{Arikan_09}. As the code length $N$ increases,
 the input channels are transformed into bit-channels that tend to become either noiseless or completely noisy channels under \emph{successive cancellation (SC) decoding}. Remarkably, the fraction of the noiseless bit-channels approaches the symmetric capacity of the transmission channel, which is referred to as \textit{channel polarization} \cite{Arikan_09}.   The theoretical capacity achieving properties of polar codes  were established by Ar{\i}kan \cite{Arikan_09, arikan2009rate}.
Recursive codes, such as the optimized codes for bitwise multistage decoding \cite{stolte2002Phd}, have been shown to be similar to polar codes \cite{el2016binary}.
Polar codes have the advantage of low encoding and SC decoding complexities of $O(N \log_2 N)$. List decoding algorithms with complexity $O(L N \log_2 N)$ for a list size $L$, have shown considerable  gains in the error correction performance of polar and recursive codes \cite{stolte2002Phd, schnabl1995soft, dumer2000recursive, list_2011}.

There have been	several attempts to construct polar codes with different rates or lengths, although not necessarily HARQ rate-compatible. Whereas the block lengths of polar codes  were restricted to powers of two \cite{Arikan_09}, polar codes with arbitrary lengths can be constructed by replacing the polarization kernel matrix $F$ with a non-singular binary matrix of size $\ell\times \ell$  \cite{Korada_Sasoglu_Urbanke_10}.
However, adjusting the polar code block length through changing the size of the kernel matrix implies different pairs of encoders and decoders for each block length, and suffers from increased decoder complexity at larger $\ell$.
 Puncturing polar codes to obtain length-compatible codes has been viewed as a process of reduction of the size of the polarization  transformation matrix $F^{\otimes n}$, such that
 exhaustive search is done at each code rate to choose the reduced matrix with the largest polarizing exponent \cite{matrix_reduction_polar}.
Changing the code rate through information shortening and random  puncturing of the output bits \cite{first_com_polar},
as well as by quasi-uniform puncturing and repetition of information bits have also been investigated \cite{harq_polar}.
Another approach used exhaustive search at each code length and desired code rate to select the puncturing pattern that minimizes the error probability
\cite{punc_pattern_polar}, although the resulting code families were not necessarily rate-compatible for HARQ  transmissions.

In this paper, we extend our previous results \cite{HARQ_RC_polar} and present a practical system for combined rate matching and bit-mapping of polar codes for HARQ rate-compatible transmissions on bit-interleaved coded modulation (BICM) channels. We propose an efficient progressive puncturing algorithm that finds the best order for puncturing the output bits on a base polar code of short length, while guaranteeing that the codes with different rates are nested and have the same information set.
Our simulation results show that this scheme achieves a decoding performance similar to that of the optimal puncturing pattern selected for a given rate and block length.
We show how a regular puncturing pattern of a longer polar code can be derived from the puncturing pattern of the base polar code, while preserving the code polarization, based on the compound polar code construction \cite{compound_polar, PolarBICM}.
This reduces the complexity of the code design, where we do not need to exhaustively search for the puncturing pattern at each desired code rate and code length.
 Moreover, the time and space complexities of operation at both the transmitter and the receiver are reduced.
 With the puncturing pattern found by the proposed progressive puncturing algorithm, we propose a simple rate matching scheme which implements a two-step compound polarization  and allows for flexible puncturing or repetition of code bits. Our scheme supports both HARQ Chase combining, where the bits of the retransmissions are the same as those of the first transmission, and HARQ incremental redundancy, where the retransmissions constitute some coded (redundancy) bits that have not been transmitted before, and may also constitute some of the previously transmitted bits. The proposed rate matching structure also integrates channel interleaving and bit-mapping for bit-interleaved coded modulation (BICM) while preserving the code polarization. Although not addressed in this paper, our proposed scheme can be applied with other polar code constructions such as the reduced-complexity relaxed polar code constructions \cite{relaxed_polar_code} or concatenated constructions that have improved burst-error correction capability \cite{mahdavifar2014performance}.

The rest of the paper is organized as follows. Preliminaries of polar codes and compound polar codes for BICM channels are reviewed in Section \ref{Pre}.
In Section \ref{Sec_rc_polar}, we describe the different aspects for the design of the proposed rate-compatible polar codes.
Theoretical analysis of the achievable rates and numerical analysis of the decoding performances of the proposed rate-compatible polar codes with different HARQ schemes are presented in Section \ref{Sec_simulation}.
Conclusions are made in Section \ref{conclude}.

\section{Preliminaries} \label{Pre}
In this section, we establish the basic notation and review the main results needed for this paper.

\subsection{Binary polar codes}
For a binary polar code of length $N=2^n$ bits, the polar encoding of an input vector is done by the polarization transformation matrix
 $F^{\otimes n}$ which is the $n$-th Kronecker power of the $2\times 2$ kernel matrix
\begin{equation} \label{Eq_G}
F=\left[\begin{array}{cc}
1 & 0 \\
1 & 1 \end{array}\right].
\end{equation}
Assume the input sequence is denoted as $u_1^N=(u_1,u_2,\ldots,u_N)$ and the corresponding codeword is denoted as $x_1^N=(x_1,x_2,\ldots,x_N)$, where $x_i$, $u_i\in\{0,1\}$ for $1\leq i\leq N$.
The encoding of a polar code can be described by $x_1^N=u_1^NB_NF_2^{\otimes n},$ where $B_N$ is a bit-reversal permutation matrix.
The coded sequence $x_1^N$ is transmitted through $N$ independent copies of a binary input discrete memoryless channel $W$ and the received sequence is denoted as $y_1^N=(y_1,y_2,\ldots,y_N)$.
The channel $W_N$ between $u_1^N$ and $y_1^N$ is described by the transitional probabilities
\begin{equation}
W_N(y_1^N|u_1^N) \triangleq P(y_1^N|x_1^N)=\prod_{i=1}^N W(y_i|x_i),
\end{equation}
where $P(\cdot|\cdot)$ represents the conditional probability.
Through  recursive \emph{channel splitting} and \emph{channel combining} operations, a bit-channel for each input bit $u_i$ under successive cancellation decoding, for $1\leq i\leq N$, is defined as
\begin{equation}
\begin{split}
W_N^{(i)}(y_1^N, u_1^{i-1}|u_i)
\overset{\triangle}{=} P(y_1^N, u_1^{i-1}|u_i).
\\
=\hspace{-5mm}\sum_{u_{i+1}^N\in{\{0,1\}}^{N-i}}\frac{1}{2^{N-1}}W_N(y_1^N|u_1^N).
\end{split}\label{eq_bitchannel}
\end{equation}
For code rate $r=k/N \leq 1$, $k$ bits of the input sequence $u_1^N$ carry information, and the rest are frozen to a known value such as zero. Hence, the decoding performance under successive cancellation decoding depends on the bit-channel reliabilities and the assignment of information bits to bit-channels.
The reliablity of bit-channels $W_N^{(i)}$ can be measured accurately on erasure channels by the Bhattacharyya parameters  $Z(W_N^{(i)})$ \cite{kailath1967divergence}, where the Bhattacharyya parameter of a binary-input channel $W$ is defined as
\begin{equation}
Z(W)\overset{\triangle}{=}\sum_{y \in \cal{Y}}\sqrt{W(y|0)W(y|1)} .
\end{equation}
For any constant $\beta <1/2$, let the subset $\cal{G}_N(W,\beta)$ of good indices be defined as
\begin{equation}
{\cal G}_N(W,\beta)\overset{\triangle}{=}\{ i: Z(W_N^{(i)})<2^{-N^\beta}/N , 1\leq i\leq N\}.
\end{equation}
Ar{\i}kan proved that polar codes are capacity achieving and showed that $\cal{G}_N(W,\beta)$ has the following property:

\begin{theorem} \label{thm_channel_polarization}\cite{Arikan_09} \cite{arikan2009rate}
For any constant $\beta < 1/2$,
\begin{equation}
\lim_{N\rightarrow \infty} \frac{|{\cal G}_N(W,\beta)|}{N}=I(W),
\end{equation}
where $I(W)$ denotes the symmetric capacity of the underlying channel $W$.
\end{theorem}

Hence, as $N$ approaches infinity, Theorem \ref{thm_channel_polarization} suggests a capacity-achieving scheme where the information bits are transmitted through the bit-channels $W_N^{(i)}$ with $i\in {\cal G}_N(W,\beta)$ . Moreover, the frame error rate (FER) of such polar code construction is shown to be less than $2^{-N^\beta}$ \cite{Arikan_09}, \cite{arikan2009rate}.

\subsection{Compound Polar Codes}
\begin{figure}[h!]
\begin{center}
\includegraphics[width= 0.35 \textwidth ]{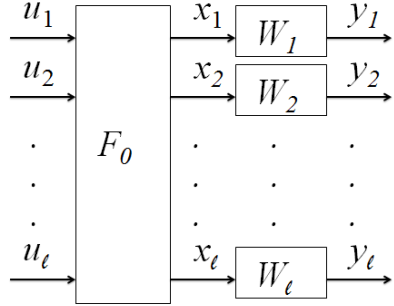}
\end{center}
\caption{Compound polar code building block for the general case of $\ell$-multi-channels.
 \label{l_multi}}
\vspace{-0.2cm}
\end{figure}

Compound polar codes \cite{compound_polar, PolarBICM} were shown to achieve the capacity of multi-channels  which constitute of multiple binary discrete memoryless channels (B-DMCs).
Hence, they were used to show a capacity acheiving scheme for transmissions on bit-interleaved coded modulation (BICM) channels \cite{PolarBICM}.
Let $W_i:{\cal X} \rightarrow {\cal Y}_i$, for $i=1,2,\ldots,\ell$, denote $\ell$ given B-DMCs and
$F_0$ be an $\ell \times \ell$ matrix satisfying the channel polarization criteria \cite{Korada_Thesis_09}, \cite{Korada_Sasoglu_Urbanke_10}.
The compound polar code construction of \cite{compound_polar} assumed a polar code of length $N=\ell2^n$, $n\geq 0$, for transmission on $\ell$ given B-DMCs.
For that, first consider a linear transformation building block depicted in Fig. \ref{l_multi}, where $x_1^\ell=u_1^\ell F_0$ and $x_i$ is transmitted through $W_i$, for $1\leq i \leq \ell.$

The compound polar code scheme can be described by deploying the linear transformation based on $F^{\otimes n}$, the $n$-th Kronecker product of  $F$ in (\ref{Eq_G}), on top of the above building block. 
The corresponding transformation is characterized by $F_C=F_0 \otimes F^{\otimes n}$, i.e, $x_1^{\ell2^n}=u_1^{\ell2^n} (F_0 \otimes F^{\otimes n})$.
 Assume that $u_i$'s are independent uniform binary random variables, and consider the compound scheme of length $N=\ell 2^n$.
Let $x_1^N=u_1^NF_{C}$ and $\tilde{W}_N$ denote the channel between $u_1^N$ and $y_1^N$ with the transition probability given by
\begin{equation}
\tilde{W}_N(y_1^N|u_1^N)=\prod_{i=1}^\ell \prod_{j=1}^{2^n}W_i(y_{(j-1)\ell+i}|x_{(j-1)\ell+i}).
\end{equation}
Similar to (\ref{eq_bitchannel}), for each information bit $u_i$, $1\leq i\leq N$, we can define the $i$th bit-channel, denoted as $(W_1\cdot W_2\cdots W_{\ell})_N^{(i)}$, by the transition probability
\begin{equation}\label{eq_com_bit_channel}
(W_1\cdot W_2\cdots W_{\ell})_N^{(i)}(y_1^N, u_1^{i-1}|u_i)\overset{\triangle}{=}\frac{1}{2^{N-1}}\sum_{\overline{u}\in\{0,1\}^{N-i}}\tilde{W}(y_1^N|(u_1^{i-1},u_i,\overline{u})).
\end{equation}
The channel polarization phenomenon also holds for compound polar codes. Given a constant $\beta <1/2$,
the subset of good channel indices $i$, ${\cal G}_N(W_1,W_2,\ldots,W_{\ell},\beta)$, defined as
\begin{equation}
{\cal G}_N(W_1,W_2,\ldots,W_{\ell},\beta)\overset{\triangle}{=}\{ i: Z((W_1\cdot W_2\cdots W_{\ell})_N^{(i)})<2^{-N^\beta}/N , 1\leq i\leq N\},
\label{compound_inf}
\end{equation}
can be shown to have the following property.
\begin{theorem}(Theorem $7$ \cite{PolarBICM})\label{thm_comp_1}
For any $\ell$ binary symmetric memoryless channel $W_1, W_2 \ldots, W_{\ell}$ and any constant $\beta <1/2$, we have

\begin{equation}
\lim_{N\rightarrow \infty}\frac{|{\cal G}_N(W_1,W_2,\ldots,W_{\ell},\beta)|}{N}=\frac{1}{\ell}\sum_{i=1}^{\ell}I(W_i),
\end{equation}
where $I(W)$ denotes the symmetric capacity of channel $W$.
\end{theorem}
The decoding of compound polar codes can be done by SC decoding  of $\ell$ polar codes each of length $N/\ell$, where the likelihoods of the bit-channels are initialized by decoding the $\ell \times \ell$ building block, which can be done with SC decoding as well if $\ell$ is power of $2$, or otherwise by maximum likelihood decoding. With such decoding, the capacity-achieving property of compound polar codes has been established in the following theorem.

\begin{theorem}(Theorem $8$ \cite{PolarBICM})\label{thm_comp_2}
For any $\beta <1/2$ and any $\ell$ binary symmetric memoryless channels $W_1, W_2 \ldots, W_{\ell}$, the rate of the compound polar code of length $N$ and information set of indices in ${\cal G}_N(W_1,W_2,\ldots,W_{\ell},\beta)$  approaches the average of the capacities of $W_1,W_2,\ldots,W_{\ell}.$ Furthermore, the probability of frame error under SC decoding is less than $2^{-N^{\beta}}.$

\end{theorem}

In this paper, we will use compound polar codes to show that our construction for punctured polar codes is capacity achieving with either binary phase shift keying (BPSK) modulation or bit-interleaved coded modulation with higher order modulations. However, our construction of compound polar codes is implicit in our proposed rate matching and bit-mapping scheme.   Furthermore, our construction does not increase the  encoding or decoding complexities compared to binary polar codes, since we show a slightly different construction where the compound code length is still a power of $2$, regardless of $\ell$.

\section{Proposed rate-compatible polar codes}\label{Sec_rc_polar}

In this section, we describe different aspects of the proposed HARQ rate-compatible polar code design.
Our goal is to design a family of nested RC codes, with rates $k/N$ or higher as in $\{ k/N, k/(N-1), \ldots, k/(N-m)\}$ by puncturing coded bits, as well as with rates
 below $k/N$ by repeating some of the coded bits at the same transmission.
Also, the transmission rate can be allowed to go above $1$ by transmitting less than $k$ bits if $(N-m) < k$.
A simple rate matching scheme together with particular bit-mappings for BICM channels is proposed, and it is shown that it can be inherently adopted to HARQ communications.
The proposed design can be deployed with systematic or non-systematic  encoding.

We begin with an efficient algorithm to obtain the puncturing order of a base polar code, which will be extended to a regular puncturing pattern of a mother polar code under the compound polarization structure.

\subsection{Progressive puncturing algorithm}
Assume an $(N,k)$ mother polar code of rate $k/N$, where $k$ is the cardinality of the information set $\mathcal{I}$.
 The set $\mathcal{I}$ can be chosen to constitute of the $k$ bit-channels with the lowest estimated bit-channel error probabilities
to minimize the union bound on the decoder block error rate (BLER)  \cite{Arikan_09},
\begin{equation}
P(\mathcal{E}) \leq \sum_{i \in \mathcal{I}} P(\mathcal{E}_i), \label{eq:PE}
\end{equation}
where $\mathcal{E}$ is the event of an error in decoding one received word, and $\mathcal{E}_i$ is the event of an error in the $i$th bit decoding.

Let the desired punctured code rate be $k/(N-m)$. The exhaustive search scheme \cite{punc_pattern_polar} considers all
possible $N \choose m$ possible puncturing patterns to puncture $m$ bits. Hence, for each possible puncturing pattern, it estimates the information set error probability (EP) (\ref{eq:PE}) after puncturing $m$ bits, and chooses the one that minimizes (\ref{eq:PE}). Alternatively, for different applications, the design criterion (\ref{eq:PE}) can be changed. Whereas exhaustive search will find the optimal puncturing pattern for that rate, it suffers from the $N \choose m$ search complexity, and does not guarantee that the  puncturing patterns, which are chosen for the different rates, are nested.

To address this, we propose an efficient greedy algorithm, referred to as \emph{progressive puncturing algorithm} (PPA), to find the best puncturing patterns for RC polar codes.
The proposed algorithm PPA finds the puncturing pattern that punctures $m+1$ bits such that the resultant puncturing pattern minimizes the design criteria (\ref{eq:PE}) and is constrained to contain the $m$ previously punctured bits found by PPA under the same information set.
Hence, it leads to a nested family of codes and supports incremental redundancy in the increments of $1$ bit, where the code of rate $k/(N-m+1)$ is obtained from the code of rate $k/(N-m)$ by transmitting the $m$th punctured bit.
The set of indices of the punctured bits is denoted by $\cP$ which is also referred to as the puncturing pattern.

To analyze the complexity of PPA, we consider a given puncturing pattern $\cP$ of size $m$.
The $(m+1)$th punctured bit is selected from the remaining $N-m$ non-punctured coded bits by testing the design criterion $N-m$ times. Hence, to find the puncturing order for all $N$ coded bits, the design criterion needs to be tested $N(N+1)/2$ for the PPA, which is significantly less than the $2^N$ searches  required with exhaustive search.

To construct a RC family from a mother code of length $N$ satisfying HARQ requirements, the information set is chosen to minimize the block error rate (BLER) of the code in the RC family with the highest code  rate  of interest, and is then fixed  across all codes with different rates in the same family. Besides having the best performance at the selected high code rate of rate less than 1, selecting the information set at the high-rate code guarantees that no zero-capacity channels are used to transmit the information bits of the lower-rate codes.  This is due to the observation that the number of zero-capacity bit-channels is equal to the number of punctured output bits, which follows from the following Lemma \cite{punc_pattern_polar}.
 \begin{lemma} \label{Lemma_punct}
  Any puncturing pattern with $m$ punctured output bits will lead to exactly $m$ bit channels with zero-capacity.
  \label{mm_lemma}
 \end{lemma}

\begin{figure}[htpb]
\begin{center}
\includegraphics[width= 0.4 \textwidth ]{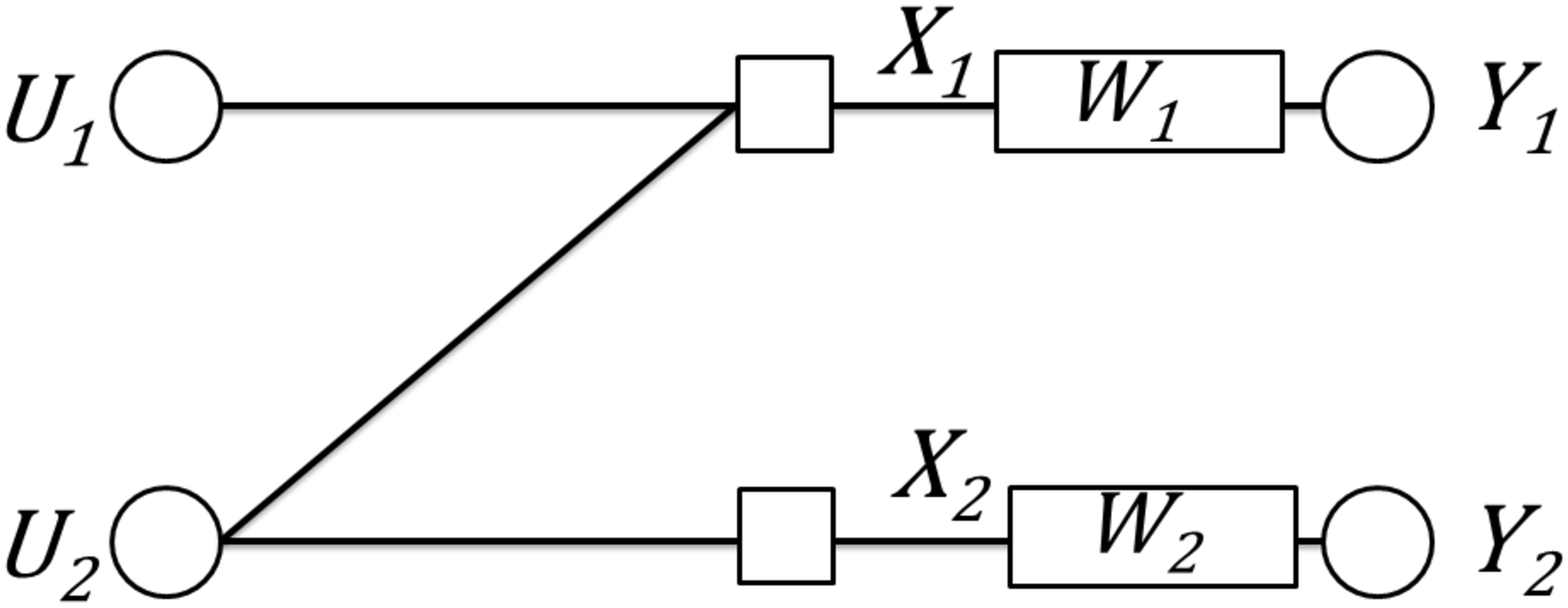}
\end{center}
\caption{Basic encoding and decoding structure of Polar codes
 \label{basic}}
\vspace{-0.2cm}
\end{figure}

 An intuitive perspective for Lemma \ref{Lemma_punct} can be seen from the basic polar encoding structure in Fig.\ref{basic}. Recall that the capacity of bit channel corresponding to $U_1$, $I(W_2^{(1)})$, is $I(U_1;Y_1,Y_2)$. If we puncture $X_1$ only, $Y_1$ is considered as noise only, then $I(W_2^{(1)})$ can be simplified to $I(U_1;Y_2)$. However, $Y_2$ is independent to $U_1$ and therefore we have $I(W_2^{(1)})=0$. Based on the sum capacity observation $I(W_2^{(1)})+I(W_2^{(2)})=I(W_1)+I(W_2)$, we can obtain $I(W_2^{(2)})=I(W_1)+I(W_2)-I(W_2^{(1)})=I(W_2)$. 
 If we puncture $X_2$ only, $Y_2$ is considered as noise only, then $I(W_2^{(1)})$ can be simplified to $I(U_1;Y_1)$. 
 Then, $U_2$ can be considered as an additive noise with the same Bernoulli distribution as $U_1$, and hence $U_1$ and $Y_1$ are independent. Therefore, $I(W_2^{(1)})=I(U_1;Y_1)=0$. Based on the sum capacity observation, $I(W_2^{(2)})=I(W_1)+I(W_2)-I(W_2^{(1)})=I(W_1)$. If we puncture $X_1$ and $X_2$, then we have $I(W_2^{(1)})=I(W_2^{(2)})=0$. Hence, puncturing any one bit at the output, results in exactly one input channel with zero capacity. Also, puncturing both bits at the output results in both input channels to have zero capacity.

\subsection{Error probability estimation} \label{Sec_EP}

For uniform channel input, the  error probability (EP) of the polarized bit-channels under genie-aided SC decoding on erasure channels can be recursively calculated through their Bhattacharrya parameters \cite{Arikan_09}.
For short codes and general channels, the bit-channel error probabilities can be estimated through Monte-Carlo simulations of the genie-aided SC decoder. Upper and lower bounds on the error probability can also be derived through channel upgrading and degrading operations \cite{Tal_vardy_2013}.
 In case of SC decoding on AWGN channels, the bit-channel EPs  can be approximated with reasonable accuracy by density evolution using the Guassian approximation (GA) method, assuming that the outputs at each SC decoding step are Gaussian random variables \cite{GA_polar}. Rather than tracking the parameters of the Gaussian distribution as in the GA method, the equivalent-SNR method \cite{stolte2002Phd}  tracks the equivalent signal to noise ratio (SNR) of the bit-channels under the Gaussian assumption, and its performance has been shown to be similar to that of the GA method \cite{el2016binary}.

 The Gaussian approximation approach \cite{GA_polar} is modified for the case when two channels with different reliability are polarized together \cite{punc_pattern_polar}. In contrary to normal polar codes, where multiplicities of the same channel are polarized together, such a case of polarizing different channels together will be observed in this work due to puncturing of the polar codes. The modified GA approach  is briefly described as follows.
Assume the all-zero codeword is transmitted and the variance of the AWGN channel is $\sigma^2$.
We denote the logarithmic likelihood ratio of the received $y_i$ as $L(y_i)$, which is assumed to have probability distribution ${\cal N}(\frac{2}{\sigma^2},\frac{4}{\sigma^2})$.
From the basic encoding and decoding structure of polar codes in Fig.\ref{basic}, we can
find $\textbf{E}(L(u_1))$ and $\textbf{E}(L(u_2))$ from $L(y_1)$ and $L(y_2)$ according to
\begin{equation}
\begin{split}
\textbf{E}[L(u_1)]
=\phi^{-1}(1-(1-\phi(\textbf{E}[L(y_1)]))(1-\phi(\textbf{E}[L(y_2)])))\label{ga_1} \\
\end{split}
\end{equation}
\begin{equation}
\textbf{E}[L(u_2)]
=\textbf{E}[L(y_1)]+\textbf{E}[L(y_2)],
\label{ga_2}
\end{equation}
where \textbf{E} denotes the expectation and
\[ \phi(x)= \left\{
 \begin{array}{ll}
 1-\frac{1}{\sqrt{4\pi x}} \int_{-\infty}^\infty \text{tanh}(\frac{u}{2})e^{-\frac{(u-x)^2}{4x}}du, & \mbox{if} \ x >0,\\
 1, & \mbox{if} \ x=0.
 \end{array}\right. \]
Through recursively applying (\ref{ga_1}) and (\ref{ga_2}) on the basic  encoding and decoding structure of a polar code,  $\textbf{E}(L(u_i))$ of each information bit $u_i$ can be calculated from the received LLRs, $L(y_1),L(y_2),\ldots, L(y_N)$, for $1\leq i\leq N$.
In case an output channel is punctured, then its variance is set to infinity.

The error probability of the $i$th bit is estimated by
\begin{equation}P(\mathcal{E}_i)=Q(\sqrt{\textbf{E}[L(u_i)/2]} )).
\label{eq:PE_i}
 \end{equation}

\begin{figure}[h!]
\begin{center}
\includegraphics[width=0.7 \textwidth ]{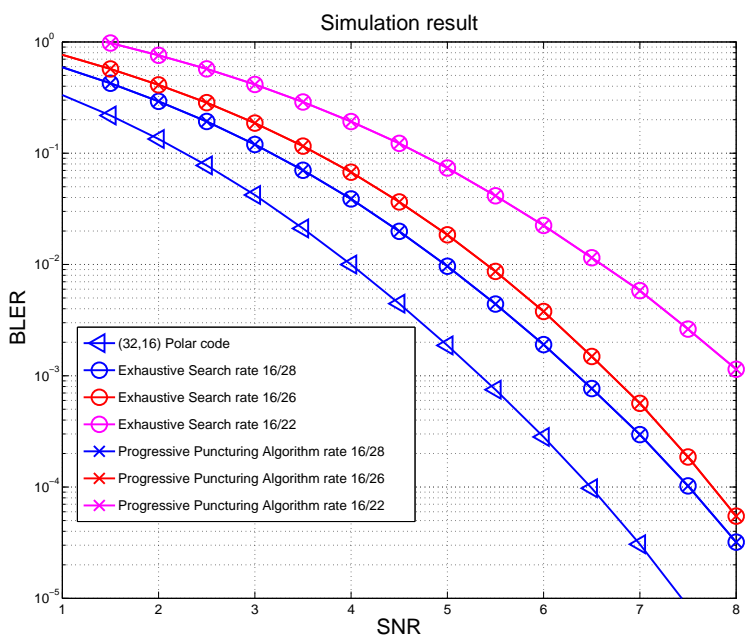}
\end{center}
\caption{Performance of progressive puncturing algorithm (PPA) versus exhaustive search puncturing of polar codes   \label{fig:PPA}}
\vspace{-0.2cm}
\end{figure}

In Fig. \ref{fig:PPA}, the decoder BLER is estimated using  (\ref{eq:PE}), where the bit error probability is found by (\ref{eq:PE_i}) over an AWGN channel after code puncturing.
 Assume a $(32,16)$ polar code designed at $\mbox{SNR}=3$ dB. The desired RC family has 4 nested codes of rates ($16/32, 16/28, 16/26, 16/22$). The optimally punctured codes are found by exhaustive search for the codes rates $\{16/28, 16/26, 16/22\}$, but are not necessarily nested. With the PPA, the nested codes are found at the desired rates. The numerical results show that the performance of the progressively punctured polar code found by PPA is shown to overlap with that of the optimally punctured polar code found by exhaustive search for each rate corresponding to $m\in \{0,4,6,10\}$.

\subsection{Two-step polarization and regular puncturing \label{Sec_regular}}

Although, the PPA has significantly smaller complexity than exhaustive search, it is not practical to repeat it for every code length and rate.
 Moreover, even the PPA algorithm becomes infeasible at very large block lengths as it requires estimating the error probabilities of all the bit-channels.
To reduce the searching complexity of PPA at large block lengths, our proposed scheme only runs the PPA on a base polar code of short length $N'$.
The base polar code is then further polarized into the polar code with the desired length $N$. This two-step polarization approach results in a regular puncturing pattern on the longer polar code of length $N$, which is derived from the puncturing sequence of the base code of length $N'$. In addition to reducing the searching complexity, another advantage of the two-step polarization approach is that it avoids storing puncturing patterns for each code length and code rate.
We explain the two-step polarization and regular puncturing in details as follows.

\begin{figure}
\begin{center}
\includegraphics[width= 0.5 \textwidth ]{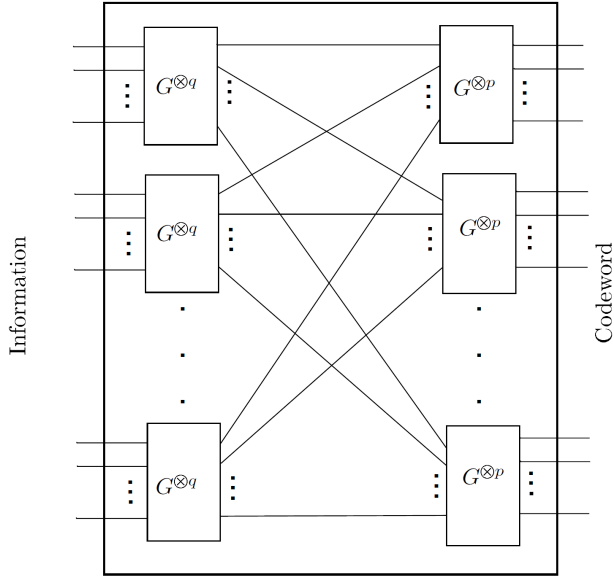}
\end{center}
\caption{Two-stage encoding scheme based on $B_{2^{n}}(F_2^{\otimes p}\otimes F_2^{\otimes q})$\label{5kron7_g}}
\vspace{-0.2cm}
\end{figure}

Consider an $(N,k)$ mother polar code of length $N=2^n$ and a base polar code of length $N'=2^p$, $0\leq p\leq n$.
The encoding structure of mother code with the generator matrix $B_{2^{n}}(F_2^{\otimes n})$ can be decomposed into two stages based on
$B_{2^{n}}(F_2^{\otimes n})=B_{2^{n}}(F_2^{\otimes p}\otimes F_2^{\otimes q})$, where $q=n-p$.
The first encoding stage consists of $2^p$ polar code encoders of length $2^q$, and the second encoding stage
consists of $2^q$ polar code encoders of length $2^p$. This is shown in Fig. \ref{5kron7_g},
where we denote the polar encoding structures $B_{2^{p}}(F_2^{\otimes p})$ and  $B_{2^q}(F_2^{\otimes q})$ by $G^{\otimes p}$ and $G^{\otimes q}$, respectively.
The PPA is run on the base code of length $2^p$ to find the progressive puncturing sequence.
To obtain a punctured code with rate of $k/(N-m2^q)$ from the mother polar code, the first $m$ bits in the progressive puncturing sequence are punctured at the output of each base polar code at the second encoding stage, i.e., at the output of each $G^{\otimes p}$ in Fig. \ref{5kron7_g}.
This results in a regular puncturing pattern on the long code, derived from base code's progressive puncturing pattern. As observed in Lemma \ref{Lemma_punct}, a punctured channel with zero capacity at the output results in exactly one input bit-channel with zero capacity. Hence, due to the regular puncturing pattern and the two stage encoding structure of Fig. \ref{5kron7_g}, all the output channels of each $G^{\otimes q}$ subcode at the first encoding stage  are transmitted over multiplicities of the same channel type, either punctured channels with zero capacity or the non-zero capacity channels.
One can observe that this construction is a generalization of the compound polar code construction   corresponding to
$2^p$ multi-channels \cite{compound_polar}. We will later show how this construction is also  beneficial when mapping the bits to symbols of higher order modulations in case of transmissions over bit-interleaved coded modulation channels.

\subsection{Circular buffer rate matching and channel interleaving}\label{Sec_interleaver}

In Section \ref{Sec_regular}, we proposed an approach to obtain a regular puncturing pattern of a large block length based on the puncturing pattern of smaller length obtained by the progressive puncturing algorithm.
However, with this regular puncturing structure, the punctured codeword length is restricted to $\ell 2^q$  and the transmission code rate is also restricted to $k/(\ell 2^q)$, for $\ell \in \{0,1,\dots, 2^p\}$.
To achieve a finer resolution of code rates by puncturing or repetition, we propose a simple rate matching scheme that preserves the regular puncturing obtained by the two step polarization and the puncturing order on the base code found by the PPA.

For a polar code of block length $N=2^n$, the proposed rate matching scheme follows the approach in Section \ref{Sec_regular} and considers the two-stage encoding structure based on $B_{2^{n}}(F_2^{\otimes p}\otimes F_2^{\otimes q})$, where $p$ and $q$ are integers satisfying $p+q=n$ and $0\leq p,q \leq n$.
First, the $2^n$ coded bits, $(x_1,x_2,\ldots,x_N)$, are arranged into a $2^{q}\times2^p$ matrix such that the $(i,j)$th element of this matrix is $x_{(i-1)2^p+j}$, for $1\leqslant i \leqslant 2^{q}$ and $1\leqslant j \leqslant 2^{p}$.
With $i=1,$ we can see that the first row of this matrix is $(x_1,x_2,\ldots,x_{2^p})$, which corresponds to the output of the first $G^{\otimes p}$ encoding structure in Fig. \ref{5kron7_g}.
Hence, in this matrix structure, each row corresponds to the output sequence of a base code encoding structure at the second encoding stage, i.e., the encoding structure $G^{\otimes p}$ in Fig. \ref{5kron7_g}.
Similarly, we can derive that each column in this matrix corresponds to the polar code encoding structure at the first encoding stage, i.e., the encoding structure $G^{\otimes q}$ in Fig. \ref{5kron7_g}.
Next, the $2^p$ columns are permuted according to the reverse of the puncturing order found by the PPA on the base code of length $2^p$. To be more specific, we rearrange the columns such that the codeword bits located in the last $m$ columns of the permuted matrix are the punctured bits of the regular puncturing pattern from two-stage encoding, which punctures the same $m$ bits in every $2^p$ bits, for $m\in\{1,2,\ldots,2^p\}$.
This is simply done by reading the first $(N-m 2^q)$ bits column-wise from the permuted matrix for a desired code rate of $k/(N-m2^q)$. This rate matching structure inherently does row-column channel interleaving to the polar codeword, where the transmitted bits are written row-wise into the matrix and read column-wise after the described column permutation.
To achieve a desired transmission rate $k/L$, the transmitted codeword bits, denoted as $(\hat{x}_1,\hat{x}_2,\ldots,\hat{x}_L)$, are read from the permuted matrix such that $\hat{x}_{(j-1)2^q+i}$ is the $(i,j)$th element of the permuted matrix. 
 If $L=N-m2^q$, then the last $m$ columns will not be transmitted, and the  two-step compound polarization with regular puncturing is preserved. Moreover, this rate matching structure does not constrain $L$  to be a multiple of $2^q$, where the first $L$ bits are always read column-wise from the permuted matrix, i.e. after the column interleaving.
 From the PPA, the column interleaving guarantees that the selected non-transmitted (punctured) bits will cause the least degradation to the BLER given by (\ref{eq:PE}).

 Moreover, for transmission at rates lower than that of the mother code where $L > N$, the matrix is treated as a circular array, where the $(N+1)$th transmitted bit is the $(1,1)$th bit of the array, and the remaining transmitted bits continue to be read column-wise from the permuted array.
This can be viewed as transmitting the non-punctured mother polar codeword, followed by transmitting a punctured polar codeword with $2N-L$ punctured bits. Since, the repetition is from the interleaved circular array, the bits selected to be punctured from the repeated codeword follows the optimized puncturing sequence. 

 According to the application, further polynomial interleaving can be applied on the output interleaved sequence. This circular array can be easily implemented in hardware as a circular buffer and has favorable implementation complexities. Its advantage is that it allows transmission at any desired rate by repetition or puncturing. Another advantage is that for rate matching or de-matching, a memory constrained device such as a wireless device, only needs to know the number of transmitted bits $L$ and the column permutation sequence, which is found by reversing the puncturing sequence of the base polar codes. Such permutation sequence will normally be specified in advance and need not be transmitted.

\begin{figure}[h!]
\begin{center}
\includegraphics[width=0.6\textwidth ]{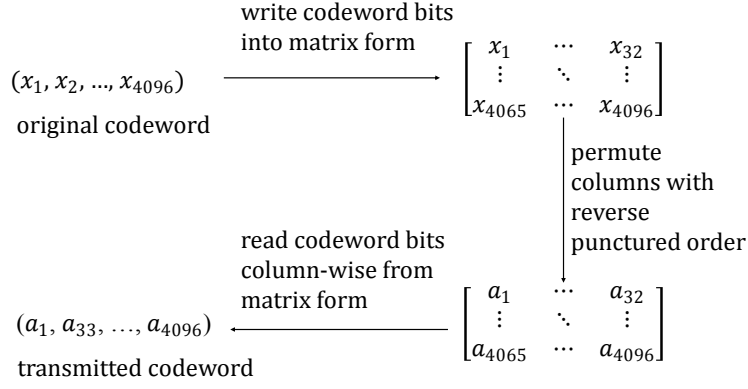}
\end{center}
\caption{Example of rearrangement of codeword of length $2^{12}$ with the circular buffer rate matching structure \label{interleaver}}
\vspace{-0.2cm}
\end{figure}

For a codeword of length $2^{12}$,  the rate matching structure is depicted in Fig. \ref{interleaver}.
In this example, we consider the encoding structure based on $B_{2^{12}}(F_2^{\otimes 5}\otimes F_2^{\otimes 7})$.
The codeword of length $2^{12}$, $(x_1,x_2,\ldots,x_{4096})$, is arranged into a $2^{7}\times2^5$ array.
Then, the columns of the array are permuted according to the reverse of the puncturing order found by applying the progressive puncturing algorithm on the based polar code of length $2^5$.
The transmitted codeword is read from this permuted array column-wise.

\begin{figure}[t]
\begin{center}
\includegraphics[width=0.8 \textwidth ]{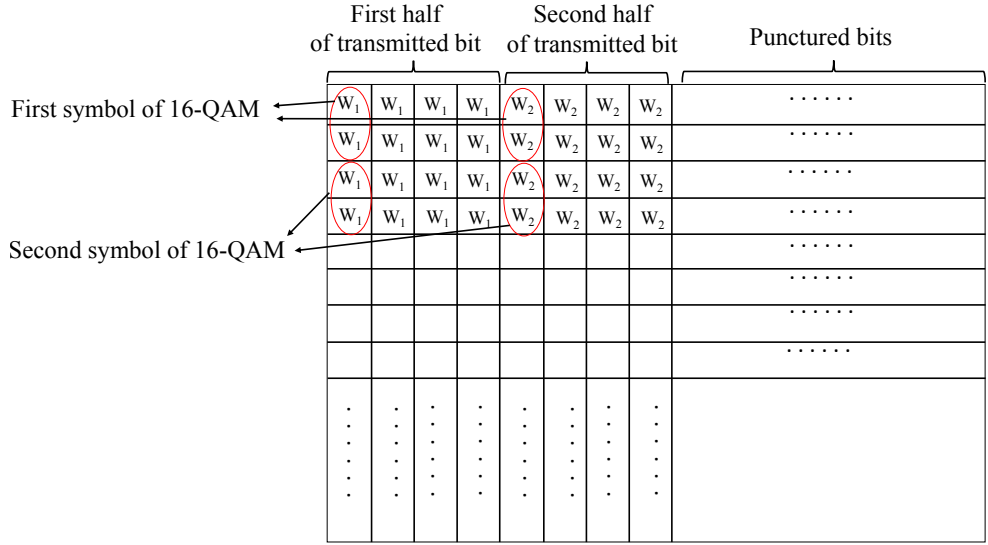}
\end{center}
\caption{Bit-mapping from coded bits to 16-QAM symbols using the circular buffer. \label{bit_mapping}}
\vspace{-0.2cm}
\end{figure}

\subsection{Bit-mapping with higher order modulations}\label{Sec_bitmapping}

 The proposed circular buffer rate matching can be combined with  bit-mapping to higher-order modulated symbols, such as 16-QAM or 64-QAM, when transmitting with
bit-interleaved coded modulation (BICM).
We apply a particular bit-mapping scheme under this rate matching structure to guarantee the polarization of the code by polarizing similar channels together, which is derived from the compound polarization structure.

For example, consider the transmission channel with $M$-QAM modulation  and gray mapping of $\log_2(M)$ bits into one symbol. This can be modeled as a multi-channel with $\log_2(M)/2$ different channel types \cite{ PolarBICM}, where BICM signal detection gives $\log_2(M)/2$ BICM channels of
different reliabilities due to symmetry of gray mapping.
The proposed bit-mapping is based on a generalized compound polar code structure, where the base polar code polarizes  $2^p$ multi-channels together, which   are mapped to the  $\log_2(M)/2$ BICM channels and the zero-capacity channels resulting from puncturing, giving $2^p$ channels at its output with different reliabilities. The $2^q$ multiplicities of each of these $2^p$ output channels are then further polarized together, with the two stage encoding structure of
  $B_{2^{n}}(F_2^{\otimes n})=B_{2^{n}}(F_2^{\otimes p}\otimes F_2^{\otimes q})$.
	
It is to be noted that the compound polar coding structure of \cite{compound_polar, PolarBICM} assumed an $l \times l$ building block for the base code for transmission over an $l$-multi-channel, and its decoding complexity is exponential in $l$ if $l$ is not a power of $2$. However, the scheme proposed here simply combines bit-mapping with the proposed  circular buffer rate matching, while keeping the decoding complexity same as that of the traditional polar code. To guarantee polarization of the resultant polar code, the idea is that each column in the circular array is assigned to only one kind of channel, whether it is a constituent BICM channel or a punctured channel. To guarantee flexibility of the combined rate matching and bit-mapping, fractional column labeling
is also allowed according to the number of punctured bits, and the number of different types of multi-channels to transmit on.
For example, the last transmitted column can be partially transmitted over a certain multi-channel and partially punctured, if the number of transmitted bits is not a multiple of $2^q$. Similarly, we can label a column for partial transmission on two multi-channels if the number of transmitted columns is not be a multiple of the number of different types of multi-channels. This gives a generalized compound polarization structure, where it allows the $2^q$ channels combined  together in the next polarization stage to be slightly different. This makes the rate matching and bit mapping design to be robust for any selected code parameters. Our numerical experiments show that these rate-matched codes have good performance, with the big advantage of preserving the low successive cancellation decoding complexities of $O(N \log N)$ for the whole code.

An example for the bit-mapping from the binary coded bits to the $16$-QAM symbols with gray mapping is shown  in Fig. \ref{bit_mapping}. The two constituent BICM channels from the $16$-QAM modulation are denoted as $W_1$ and $W_2$.
The proposed bit-mapping scheme is combined with rate matching, such that the last $m$ columns are punctured, the first half of the non-punctured bits are transmitted on the first constituent BICM channel, and the second half of the non-punctured bits are transmitted on the other constituent channel. For each $16$-QAM gray-encoded symbol, two bits experiencing the same BICM channel are mapped from the first half, and the remaining two bits are mapped from the second half. Similarly to assigning the last columns of the permuted matrix to the punctured channels which have zero capacity, one can argue that it is best to assign the columns to the bit-channels in descending order of their channel capacities, giving the first-half and second-half assignment shown in Fig. \ref{bit_mapping}. However, different assignments of the transmitted columns to the $\log_2(M)/2$ bit channels, such as an alternating column assignment, are possible.

Punctured polar coding with $64$-QAM modulation and gray mapping can be also modeled as polar coding over a multi-channel with four different channel types, three channels with different reliabilities are due to $64$-QAM BICM with gray-mapping and the fourth channel is  the zero-capacity channel due to puncturing.
After rate matching, the transmitted columns are grouped into three groups of the same size such that each group is mapped to a different BICM channel type.
The idea in this bit-to-symbol mapping is to guarantee, whenever possible, that each component $G^{\otimes q}$ polarizes over the same channel type. Hence, different assignments of the columns to the BICM bit-channels are possible.

\subsection{HARQ Chase combining and incremental redundancy}\label{Sec_HARQ}

The proposed rate matching scheme allows transmission of rate-compatible polar codes with both Chase combing (CC) and incremental redundancy (IR) types of HARQ communications. Assume a set of coded bits are selected for transmission, but the decoding result of that transmission indicated a decoding error or decoding failure. With CC, the same set of coded bits for that transmission are retransmitted. With IR, the new transmission can include some of the coded bits which have not been previously transmitted, as well as some of those previously transmitted coded bits.

The proposed rate matching and bit-mapping scheme allows seamless IR or CC HARQ transmissions.
Assume the transmission rate of $k/L$ is fixed across all HARQ transmissions and a maximum of $t$ transmissions.
With HARQ CC, all transmissions will send the $L$ bits read column-wise from the permuted matrix, starting from the first column. With IR, the $L$ bits are also transmitted column-wise at each transmission, but  starting from the $((r-1)2^p/t + 1)$th column for the $r$th transmission. The BICM bit-mapping is then applied by circularly shifting the BICM column assignments, as described in Sec. \ref{Sec_bitmapping},  for $(r-1)2^p/t$ columns.
Other combinations of the IR column assignments of each transmission with the BICM bit-channel assignments that preserve the two-step polarization structure are also possible. For example, alternating BICM column assignments which already have a cyclic pattern need not be circularly shifted.
With the proposed circular column offset between HARQ transmissions,  if all $t$ transmissions are sent with $L > N/t$, all coded bits would have been sent, while retransmitting some coded bits. Moreover, this HARQ scheme also supports IR transmissions for the case $L > N$, where the code rate of each transmission is lower than that of the mother code rate using the circular array technique. In such a case, it is still advantageous to start each transmission at the $\left((r-1)2^p/t \right)$th column for transmission diversity and allowing different bits to be repeated at each transmission.

We also note that although the described scheme for rate matching, bit-mapping, and HARQ assignments does not need the coded bits to be systematically encoded, it also works with systematic encoding. There has been multiple methods proposed for systematic encoding of polar codes \cite{stolte2002Phd, el2016binary, SystematicPolar, EfficientSystematic, sarkis2016flexible}. It is well-known for many coding schemes that systematic encoding can be done by erasure decoding,  \emph{cf.} \cite{stolte2002Phd, SystematicPolar}. Systematic encoding of polar codes
can be viewed as encoding the input vector twice \cite{sarkis2016flexible}, where  the output word after the first regular (non-systematic) encoding is re-encoded after setting its indices to zero at the frozen bit positions. Hence, decoding of systematically encoded polar codes can be done by simply decoding the second encoded word to give a non-systematic input word, which is re-encoded to get the systematic codeword.  Since puncturing is done at the output codeword, it follows that the same puncturing sequence found for non-systematic polar codes by the PPA still applies for systematic polar codes and our scheme is transparent to whether systematic or non-systematic encoding has been used. The latency and computational complexity of systematic encoding and decoding depends on the implementation \cite{sarkis2016flexible}. It also to be noted that for any coding scheme, systematic encoding does not change the BLER, although it can improve the information bit error rate (BER).

\section{Analysis of proposed RC polar codes}\label{Sec_simulation}
In this section, we further explore the proposed scheme by providing a theoretical analysis on the achievable rates followed by numerical simulation results.

\subsection{Theoretical analysis}
We first consider a base polar code of length $N'=2^p$ with the linear transformation between information sequence $u_1^{N'}$ and codeword $x_1^{N'}$ characterized by $G^{\otimes p}$ in Fig. \ref{5kron7_g}.
Assume a puncturing pattern $\cP$ of size $m$ on the base polar code.
Instead of transmitting output bits on identical channels as in a regular polar construction, the punctured bits, $x_i: i \in \cP$, are transmitted on zero capacity channels and the remaining
 $x_i$s are transmitted through $N'-m$ independent copies of a B-DMC $W$.
The above channel model can be regarded as $N'$ B-DMCs consisting of $N'-m$ independent copies of the B-DMC $W$ and $m$ zero-capacity channels.
Hence, the channel assignments together with the base polar transformation can be depicted as the building block for $\ell$-multi-channels in Fig. \ref{l_multi} with $\ell=2^p.$
Let $y_1^{N'}$ denote the channel output.
The modified channel $\tilde{W}_{N'}$, between $u_1^{N'}$ and $y_1^{N'}$, are described by the transitional probabilities
\begin{equation}
\tilde{W}_{N'}(y_1^{N'}|u_1^{N'}) \triangleq  \prod_{i \in [N']\setminus \cP} W(y_i|x_i),
\end{equation}
where $[N']$ is the set of indices $1,2,\dots,N'$ and $x_1^{N'}=u_1^{N'}G^{\otimes p}$.
Similar to (\ref{eq_com_bit_channel}), the transition probabilities of modified bit-channels for $i \in [N']$ can be defined by
\begin{equation}
\begin{split}
\tilde{W}_{N'}^{(i)}(y_1^{N'}, u_1^{i-1}|u_i)
=\hspace{-5mm}\sum_{u_{i+1}^{N'}\in{\{0,1\}}^{N'-i}}\frac{1}{2^{N'-1}}\tilde{W}_{N'}(y_1^{N'}|u_1^{N'}).
\end{split}\label{eq_bit_channel_punc_l}
\end{equation}

The capacity achieving property of the rate-matched polar codes in (\ref{eq_bit_channel_punc_l}) are found by using arguments similar to those used in showing the capacity achieving property of the compound polar codes \cite{compound_polar, PolarBICM}. Consider the building block structure in Fig. \ref{l_multi} where the $N'$ B-DMCs, $W_1$, $W_2$, $\ldots$, $W_{N'}$ constitute of $N'-m$ independent copies of a B-DMC $W$ and $m$ zero-capacity channels. Let ${I}(W)$ denotes the symmetric capacity of channel $W$. Then, from Lemma 4 \cite{PolarBICM}, we have the following lemma.
\begin{lemma}
\label{sum-capacity}
Consider the polar transformation  $G^{\otimes p}$ of an input with length $N'= 2^p$ for transmission on
 a B-DMC $W$,  then for any puncturing pattern $\cP$ applied on the output of $G^{\otimes p}$ with cardinality $m$, such that  $m < N'$,  the sum capacity is
$
\sum^{N'}_{i=1} {I}(\tilde{W}^{(i)}_{N'}) = (N' - m) {I}(W)$.

\end{lemma}

Consider the two-step polar transformation of length $N=2^n$ between the information sequence $u_1^N$ and codeword $x_1^N$ with base polarization block $G^{\otimes p}$ of length $N' = 2^p$  as shown in Fig. \ref{5kron7_g}.
Assume a puncturing pattern $\cP$ of length $m$ on the base polar code.
We consider a regular puncturing pattern $\cP_{reg}$ on $x_1^N$ defined as $\cP_{reg}=\{i: i  \ \text{mod} \ 2^p \in \cP, 1\leq i\leq N\},$ i.e., $x_i$ is transmitted on zero capacity channel if $i \in \cP_{reg}$ and the remaining
 $x_i$s are transmitted on independent copies of a B-DMC $W$.
Let $y_1^{N}$ denote the channel output with $y_i=0$ for $i \in \cP_{reg}$.
The modified channel $\tilde{W}_{N}$, between $u_1^{N}$ and $y_1^{N}$, can be described by the transitional probabilities
\begin{equation}
\tilde{W}_{N}(y_1^{N}|u_1^{N}) \triangleq  \prod_{i \in [N]\setminus \cP_{reg}} W(y_i|x_i),
\end{equation}
where $[N]$ is the set of indices $1,2,\dots,N$.
Then, we can define the modified bit-channels for $i \in [N]$ by
\begin{equation}\label{eq_bit_channel_punc}
\begin{split}
\tilde{W}_{N}^{(i)}(y_1^{N}, u_1^{i-1}|u_i)
=\hspace{-5mm}\sum_{u_{i+1}^{N}\in{\{0,1\}}^{N-i}}\frac{1}{2^{N-1}}\tilde{W}_{N}(y_1^{N}|u_1^{N}).
\end{split}
\end{equation}
For a constant $\beta <1/2$,
consider the subset of indices $i$, ${\cal G}_N(\tilde{W}_N,\beta)$, defined as
\begin{equation}
{\cal G}_N(\tilde{W}_N,\beta)\overset{\triangle}{=}\{ i: Z(\tilde{W}_N^{(i)})<2^{-N^\beta}/N , 1\leq i\leq N\}.
\label{compound_inf}
\end{equation}

The achievable rate of the proposed two-step polarization with puncturing pattern $\cP_{reg}$ can be derived based on ${\cal G}_N(\tilde{W}_N,\beta)$ in the following theorem.

\begin{theorem}\label{thm_comp_punc_1}
Consider the two-step polarization of length $N$ with a base code of length $2^p$ and the regular puncturing pattern $\cP_{reg}$ defined above.
For any constant $\beta <1/2$, we have

\begin{equation}
\lim_{N\rightarrow \infty}\frac{|{\cal G}_N(\tilde{W}_N,\beta)|}{N}= \left(1-\frac{m}{2^p}\right)\cC(W),
\end{equation}
where $\cC(W)$ denotes the symmetric capacity of the transmission channel $W$.
\end{theorem}
\beginofproof
By construction, $\cP_{reg}$ punctures the same indices at the output of each of the $2^q$ blocks representing $G^{\otimes p}$ block. Hence, the same indices of the $2^q$ multiplicities of the base code $G^{\otimes p}$ will be transmitted on zero capacity channels, and the remaining $2^p-m$ indices will be transmitted on multiplicities of $W$.  Each base code polarizes these channels together providing identical sets of transformed channels.  The two-step polarization structure of Fig. \ref{5kron7_g} guarantees that the $2^q$ multiplicities of each of the channels transformed by the base code are polarized together by a $G^{\otimes q}$ polar subcode.  By the capacity achieving property of polar codes \cite{Arikan_09},  given by Theorem  \ref{thm_channel_polarization}, the rate of each subcode approaches the capacity of each transformed channel.
By Lemma \ref{Lemma_punct}, each base code will have exactly $m$ transformed channels at its output with zero capacity, hence $m$ of the  $G^{\otimes q}$ polar subcodes will polarize zero capacity channels, and the rate of the other $2^p-m$ polar subcodes will approach the capacities of the non-zero channels. The rate of the polar code approaches the sum rate of its subcodes.  The rest of the proof follows by observing that this construction is  a  compound polar code transmitted on a $2^p$-multichannel constituting of $m$ zero capacity channels and $2^p-m$ independent copies of $W$.
By Theorem \ref{thm_comp_1}, the rate of the two-stage encoded polar code approaches the average of the capacities of $\tilde{W}^{(i)}_{2^p}$, i.e.  for any constant $\beta <1/2$
\begin{equation}\nonumber
\lim_{N\rightarrow \infty}\frac{|{\cal G}_N(\tilde{W}_N,\beta)|}{N}=\frac{1}{2^p}\sum_{i=1}^{2^p}I(\tilde{W}^{(i)}_{2^p})=\frac{2^p-m}{2^p}I(W).
\end{equation}
\endofproof

 By Theorem \ref{thm_comp_punc_1} and  Theorem \ref{thm_comp_2}, the two-step polarization of Fig. \ref{5kron7_g} is shown to result in a family of capacity-achieving punctured polar codes. The codes are shown to approach the average capacity after puncturing a fraction $m/2^p$ of the transmitted bits.

\begin{theorem}
Consider the two-stage polarization of length $N$ with a polarization base of length $N'=2^p$ of which $m$ bits are punctured.   As code length $N$ goes to infinity, and $\beta < 1/2$, there exists a family of two-step polarized codes approaching the rate $I_m=\left(1-\frac{m}{N'}\right)I(W)$ with vanishing probability of error bounded by $2^{-N^{\beta}}$ under successive cancellation decoding.
\end{theorem}

We extend this result for the case of BICM channels where the non-punctured channels are not identical but constitute of $\log_2(M)$ different channels. With the  bit-mapping assignment that follows from the column-wise assignment of different bit-channels in the circular array, it follows from the proof of Theorem
 \ref{thm_comp_punc_1}  that
 \begin{equation}
 I_m = \lim_{N\rightarrow \infty}\frac{|{\cal G}_N(\tilde{W}_N,\beta)|}{N}=\frac{1}{2^p}\sum_{i=1}^{2^p}I(\tilde{W}^{(i)}_{2^p})=\left(1-\frac{m}{2^p}\right) \textbf{E}[I(W)]
 \end{equation}
  where  $\textbf{E}[I(W)] = \frac{1}{\log_2(M)}\sum_{i=1}^{\log_2(M)} I(W_i)$ is the average symmetric capacity of the BICM channels as described in subsection \ref{Sec_interleaver}.

\subsection{Numerical analysis}
Next, we provide numerical simulations for the proposed interleaved RC polar codes with BICM.
 Let $s_k$ be the $k$th transmitted QAM symbol, then the $k$th received symbol is $r_k=a_k s_k + n_k$, where $a_k$ is the zero-mean complex Gaussian fading coefficient or the Rayleigh fading coefficient for QAM or BPSK, respectively.
 The noise $n_k$ is the additive white Gaussian noise (AWGN) with signal to noise ratio (SNR)
 $10 \log_{10}(1/(2 \sigma^2)$ in case of QAM, or $10 \log_{10}(1/\sigma^2)$ in case of BPSK.

\begin{figure}[ht!]
\begin{center}
\includegraphics[width=0.7 \textwidth ]{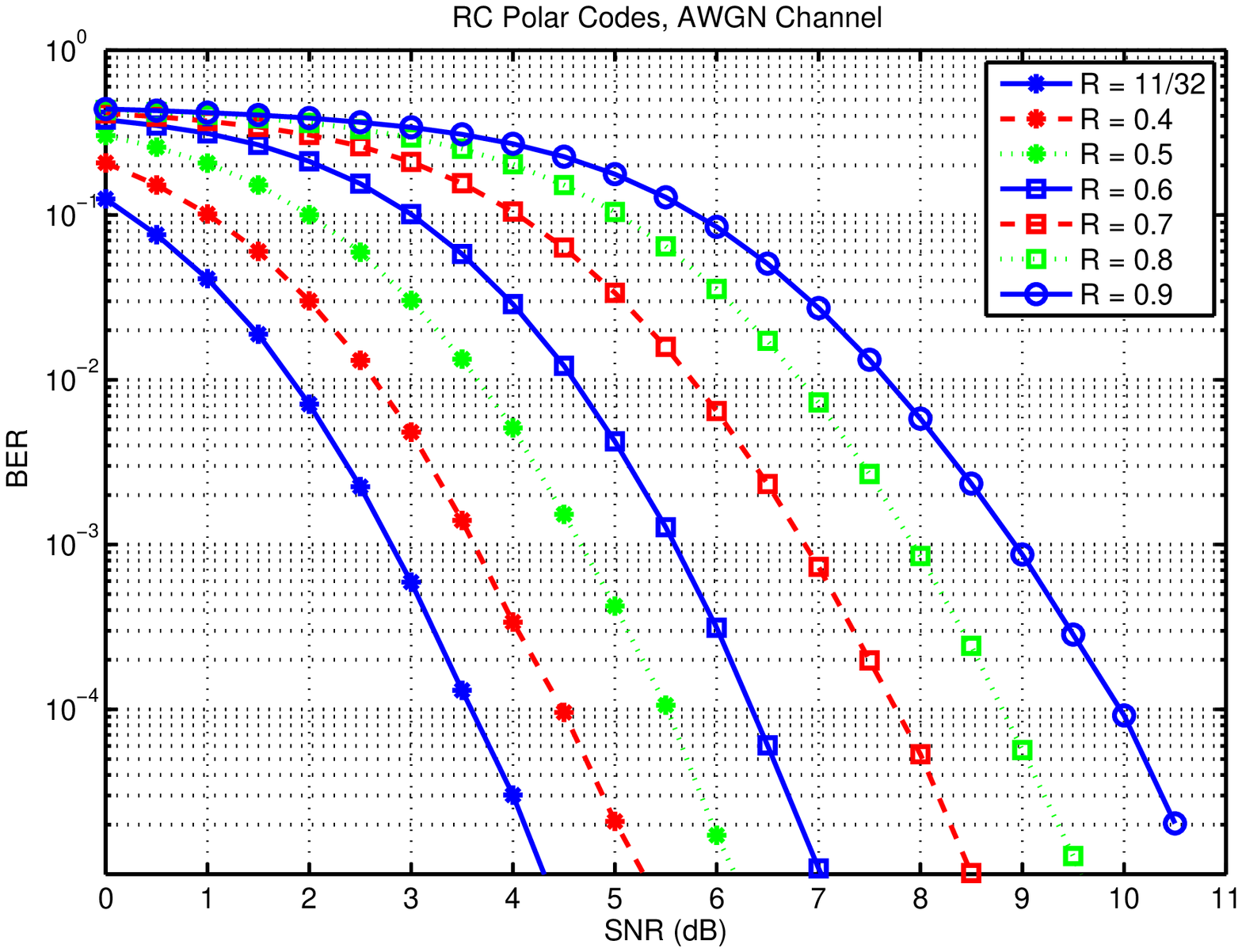}\\

\includegraphics[width=0.7 \textwidth ]{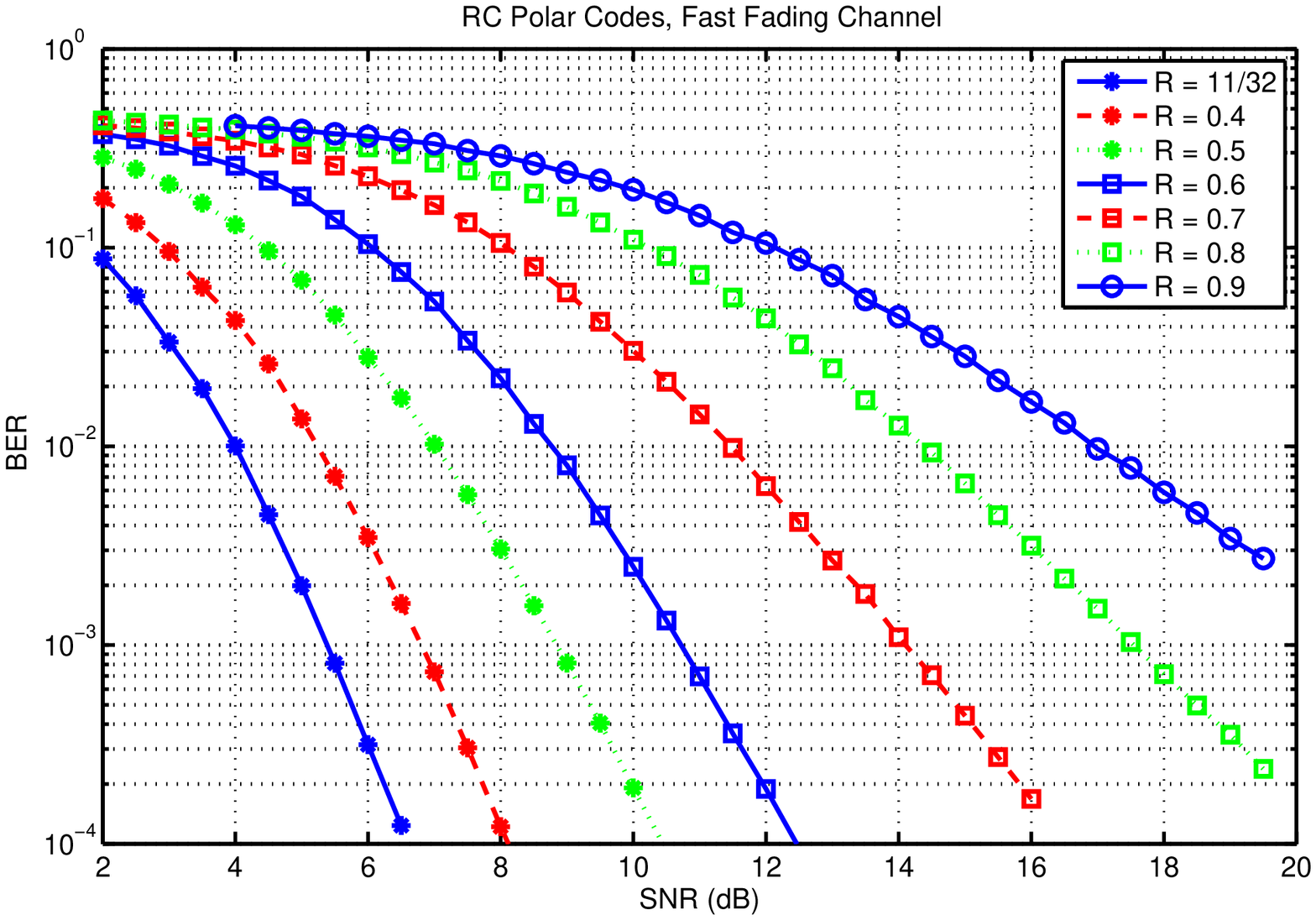}
\end{center}
\caption{BER of RC polar codes with  a fixed information set of size $88$ at different code rates under AWGN (top) and fast fading (bottom) channels \label{RC_256_fading}}
\vspace{-0.2cm}
\end{figure}

\emph{Example 1:} 
We consider an example for rate matching of short polar codes, which is suitable for adoption in the proposed 3GPP NR control channel \cite{3GPPNR}. The progressive puncturing sequence is found on a base code of length $32$, from which a RC family with a mother code of length $256$ is constructed. 
The PPA is run with the GA at $\mbox{SNR}=3.5$ dB on the base code with rate $11/32$, which gave the base code puncturing sequence of length $32$ which is simply shown as follows: $(0,\ 16,\ 8,\ 24,\ 2,\ 20,\ 26,\ 12,\ 10,\ 18,\ 4,\ 22,\ 25,\ 6,\ 13,\ 14,\
1,\ 17,\ 28,\ 3,\ 5,\ 9,\ 29,\ 11,\ 19,\ 7,\ 21,\\ 15,\ 23,\ 27,\ 30,\ 31)$.
The BER performance by SC decoding of the RC family with $N=256$ and $|\mathcal{I}|=88$ bits at code rates
 $R \in \{0.9, 0.8, 0.7, 0.6, 0.5, 0.4, 11/32\}$, is shown in  Fig. \ref{RC_256_fading}, for BPSK transmissions on both AWGN and fast fading channels. To select the information set, the bit-channel error probabilities are estimated by genie-aided SC decoding of the codes with rate $R=88/98$ at SNRs of $3.5$ dB and $6.5$ dB  on the AWGN and the fast fading channels, respectively.
\fexmpl

In the following examples, the puncturing sequence is found by applying the PPA on a base code of length $1024$, which will be utilized to construct RC families with longer code lengths.

\begin{figure}[ht!]
\begin{center}
\includegraphics[width=0.8\textwidth ]{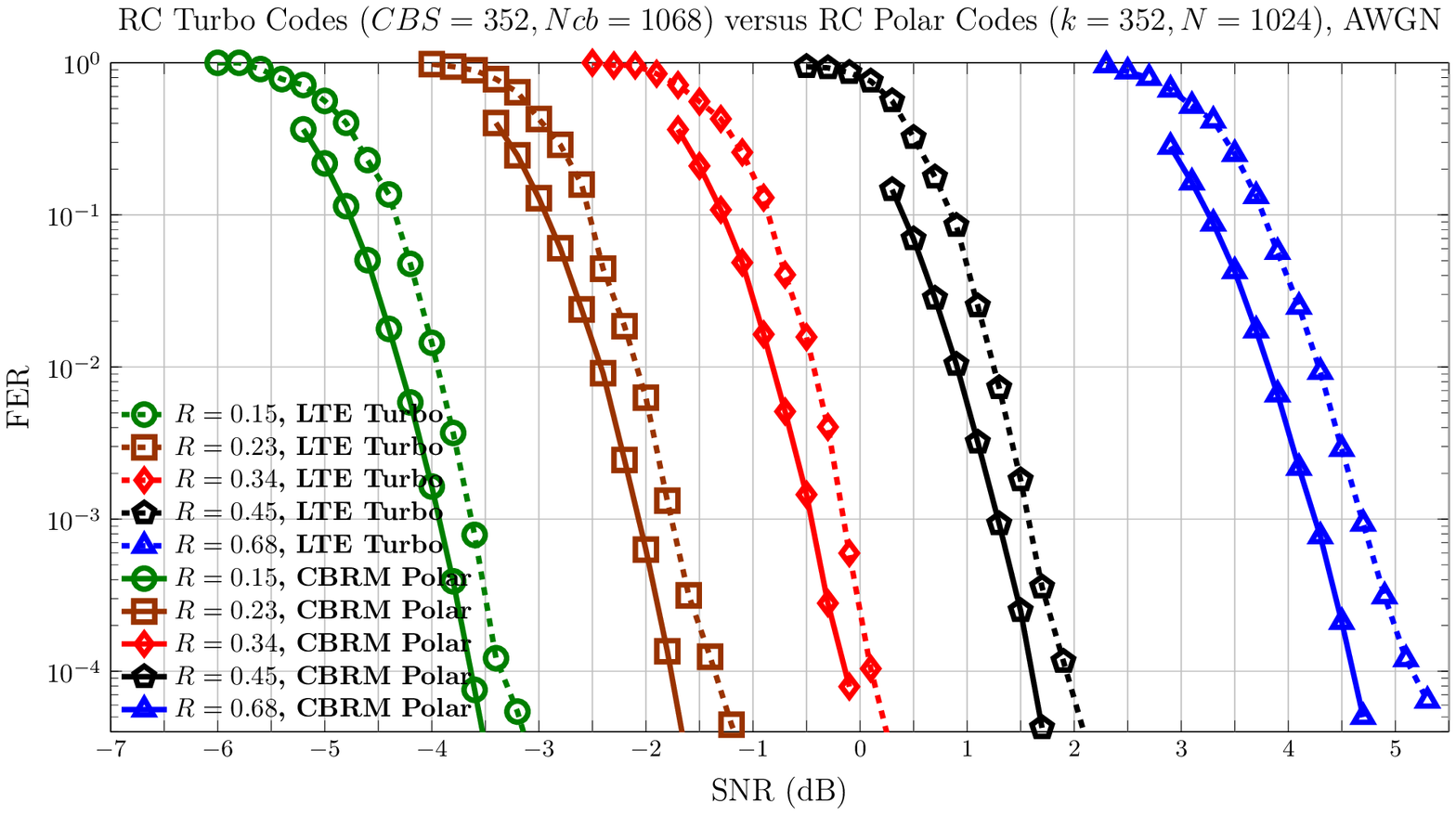}\\
\vspace{-10cm}
\includegraphics[width=0.8\textwidth ]{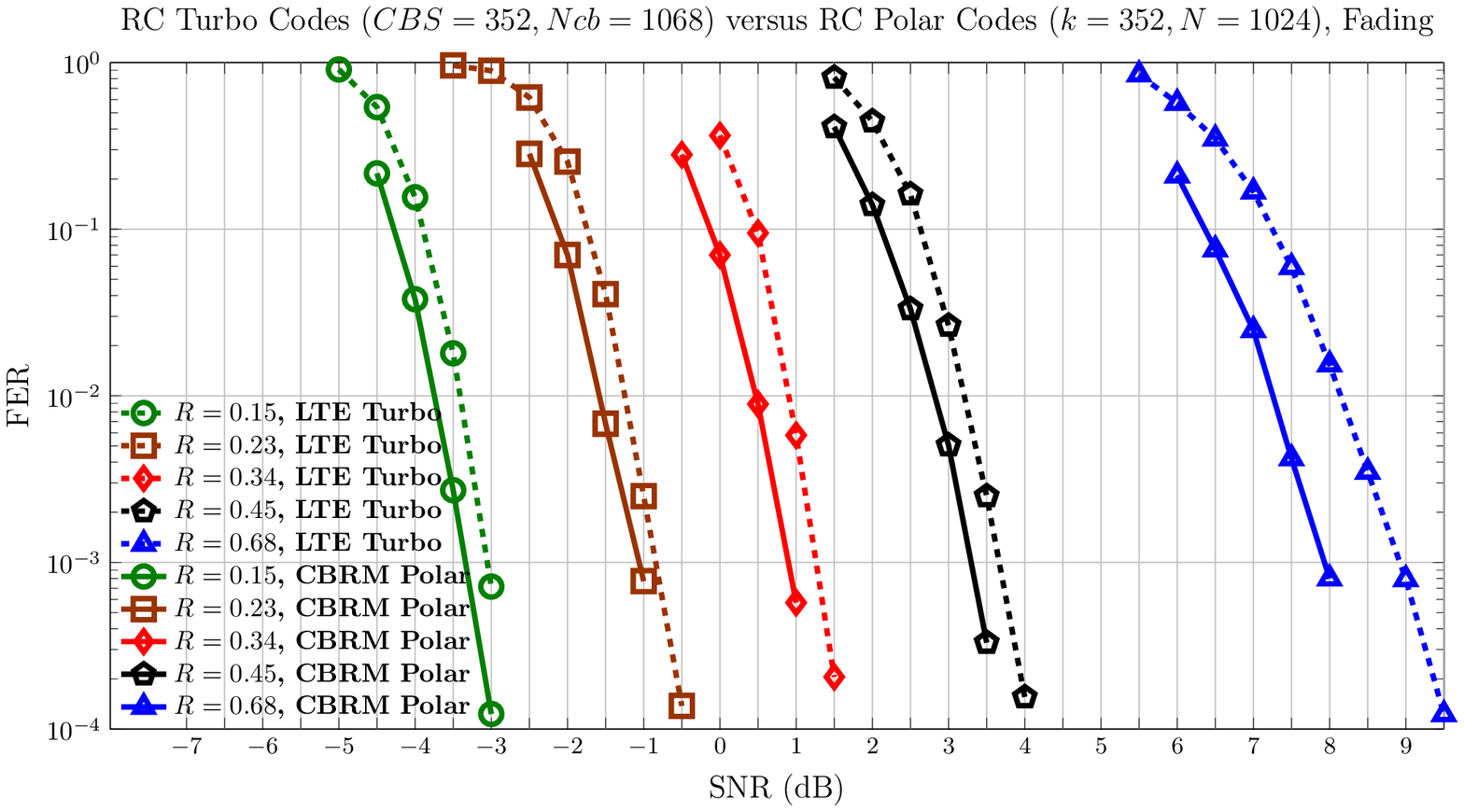}
\end{center}
\vspace{-5cm}
\caption{Comparison of LTE turbo codes with the RC circular buffer rate-matched (CBRM) polar codes, having  a fixed information set of size $352$ at different code rates under AWGN (top) and fast fading (bottom) channels \label{RC_1024_fading}}
\vspace{-0.2cm}
\end{figure}

\emph{Example 2:} We design a RC circular buffer rate-matched (CBRM) polar code family with parameters and constraints similar to that of the 3GPP LTE turbo codes. For fair comparison with the LTE turbo code, the polar code is systematically encoded. Also, a cyclic redundancy check (CRC) code with $24$ check bits is used to encode the information vector, similar to the LTE case. The mother polar code length is $N=1024$, and its information block length is $k=352$ non-frozen bits which consists of $328$ information bits and $24$ CRC bits. List SC decoding with list size of $32$ and CRC error detection, is used for polar decoding. The performance of this RC family is compared with that of LTE turbo codes with the same information block length of $352$ bits and mother code rate of $1/3$.
Also, since in LTE, the same turbo code is used for both AWGN and fading channels, we design one RC family which is tested on both the AWGN and fast fading channels. The puncturing sequence is found on the base code of length $1024$ using the PPA. The information set for the RC family is selected at the code of rate $R=0.68$,  which is the highest rate of interest in the RC family, and at a SNR of $3$ dB. To select the information set, the bit-channel error probabilities are estimated by the Gaussian approximation approach as described in Section \ref{Sec_EP}. The RC family is constructed either by puncturing if the rate is above $11/32$ or retransmitting coded bits circularly as described in Section \ref{Sec_interleaver}. The two-stage encoding structure is described by $B_{2^{10}}(F_2^{\otimes 10} \otimes F_2^{\otimes 0})$.
Fig. \ref{RC_1024_fading} shows the frame error rate (FER) performances on the RC family  at different code rates for QPSK transmissions on both AWGN and fast fading channels.
We note that the RC family shows a comparable FER performance to LTE turbo codes, with at least $0.2$ dB coding gain across all code rates, over both AWGN and fast fading channels with no signs of error floors. \fexmpl

\begin{figure}[ht!]
\begin{center}
\vspace{-3cm}
\includegraphics[width=0.8\textwidth ]{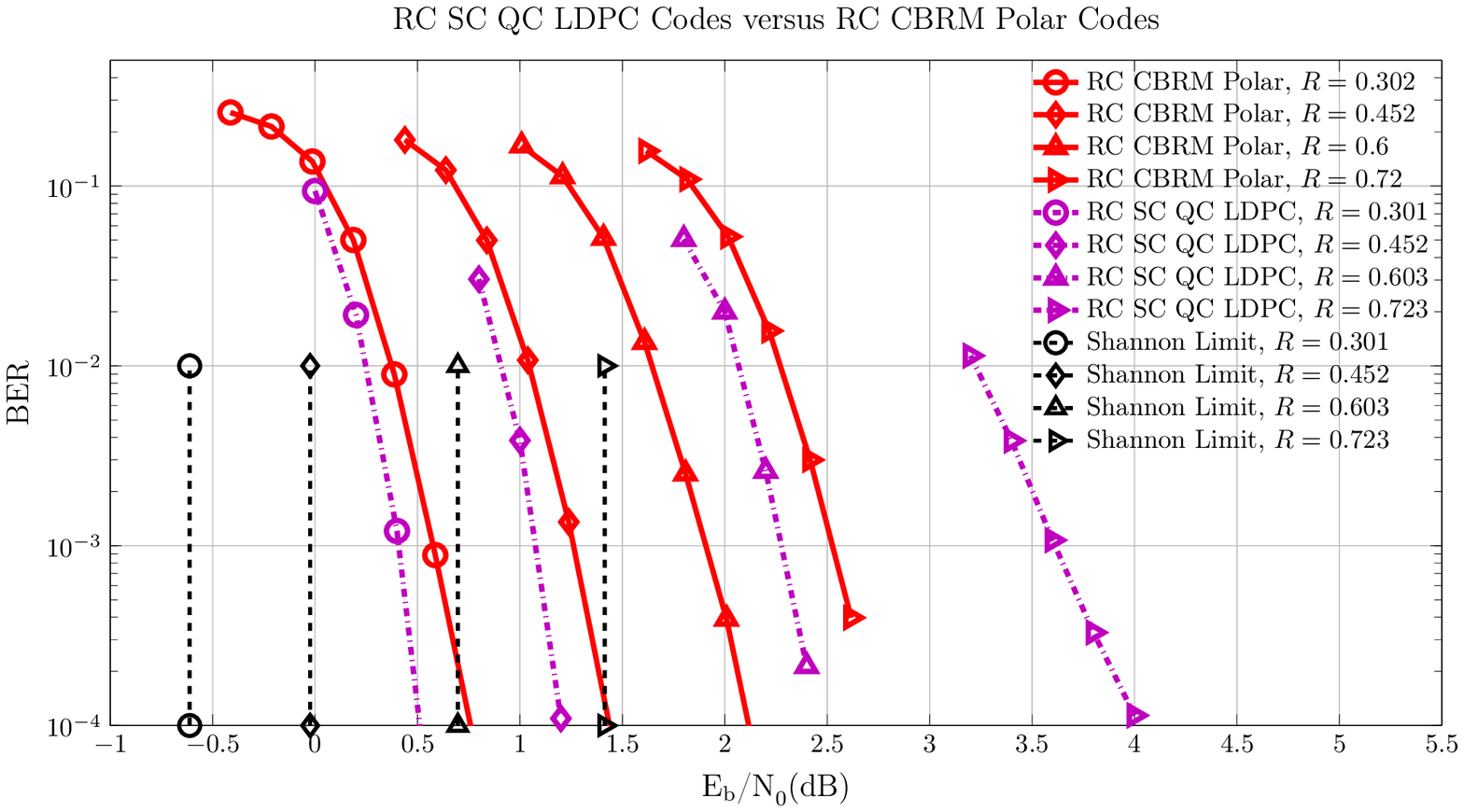}\\
\vspace{-7cm}
\includegraphics[width=0.8 \textwidth ]{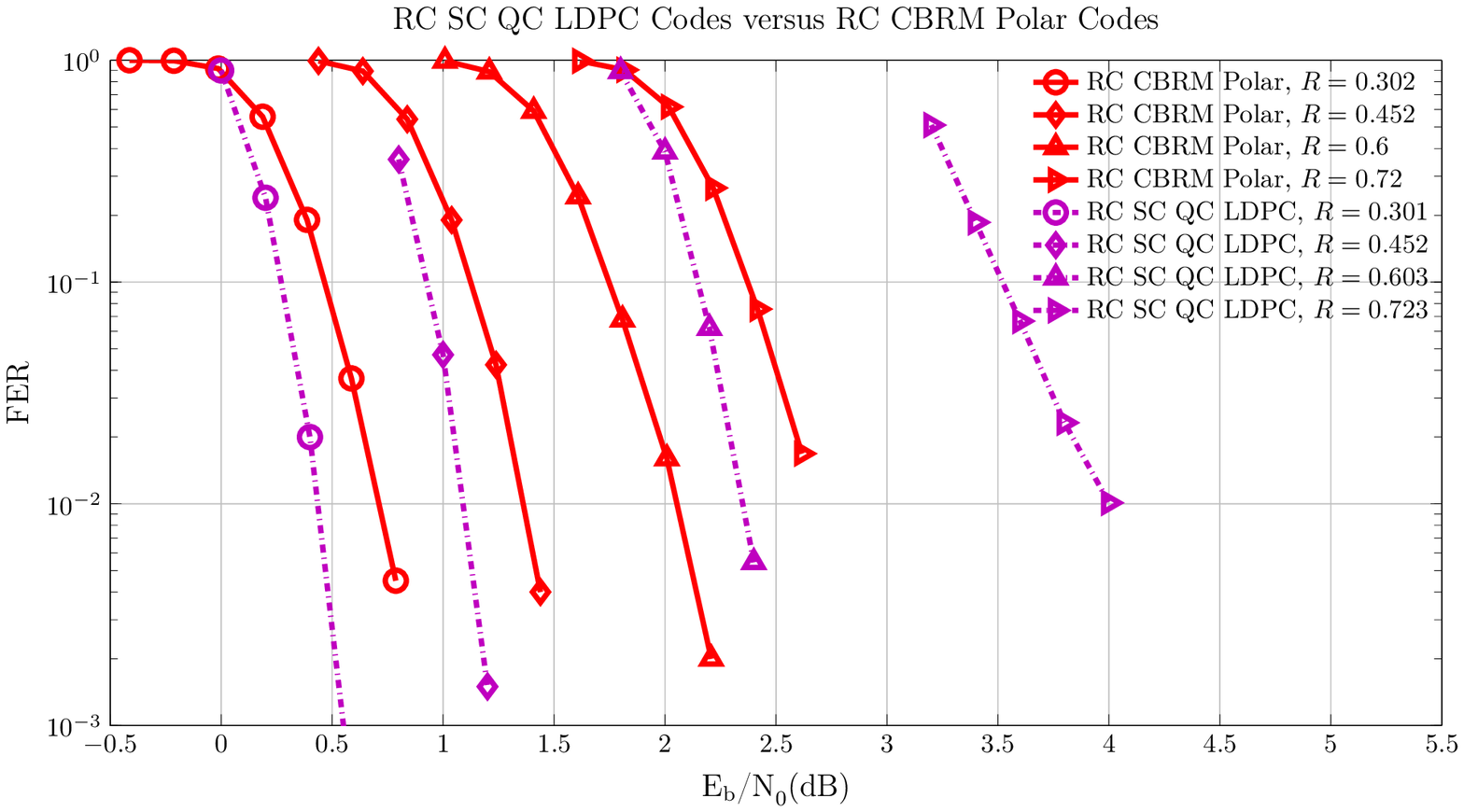}
\end{center}
\vspace{-5cm}
\caption{BER (top) and FER (bottom) of RC CBRM polar codes with a fixed information set of size $4951$ at different code rates are compared with RC SC-QC LDPC codes (\cite{SCQCLDPC} $C_2$), with BPSK on AWGN channels. \label{fig_RC_SP}}
\end{figure}

\begin{figure}[ht!]
\begin{center}
\includegraphics[width=0.8 \textwidth, angle=270, origin=c ]{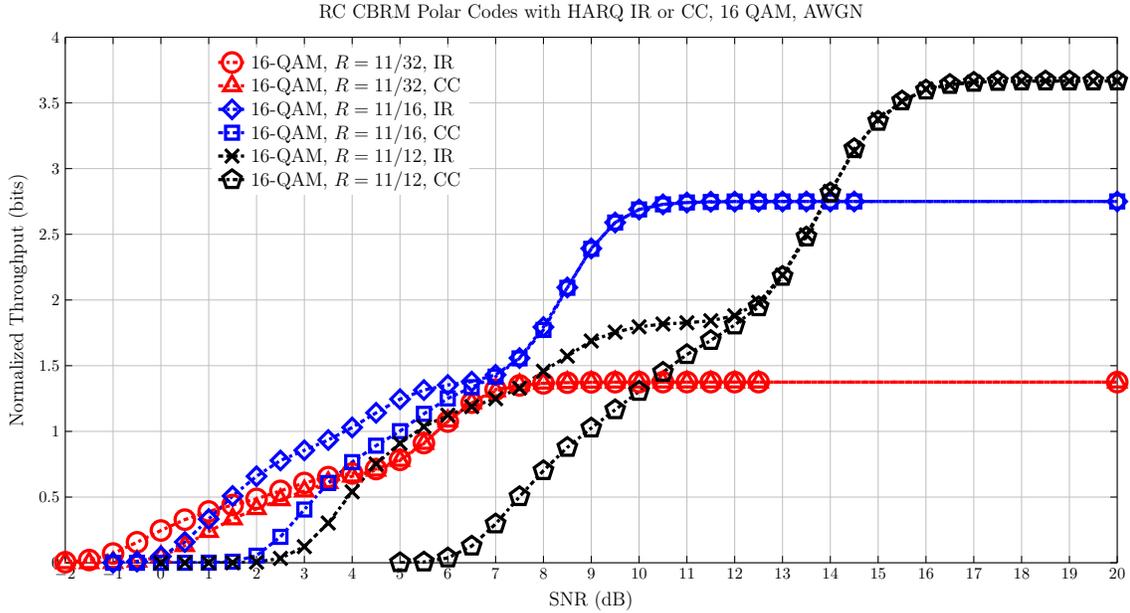}
\end{center}
\vspace{-3cm}
\caption{Comparison of CC HARQ and IR HARQ with proposed circular buffer rate matching and bit mapping, with 16-QAM on AWGN channel, $N=1024, N'=32$.}

\label{HARQ_1024_16qam}
\end{figure}

\begin{figure}[ht!]
\begin{center}
\includegraphics[width=0.8 \textwidth, angle=270, origin=c ]{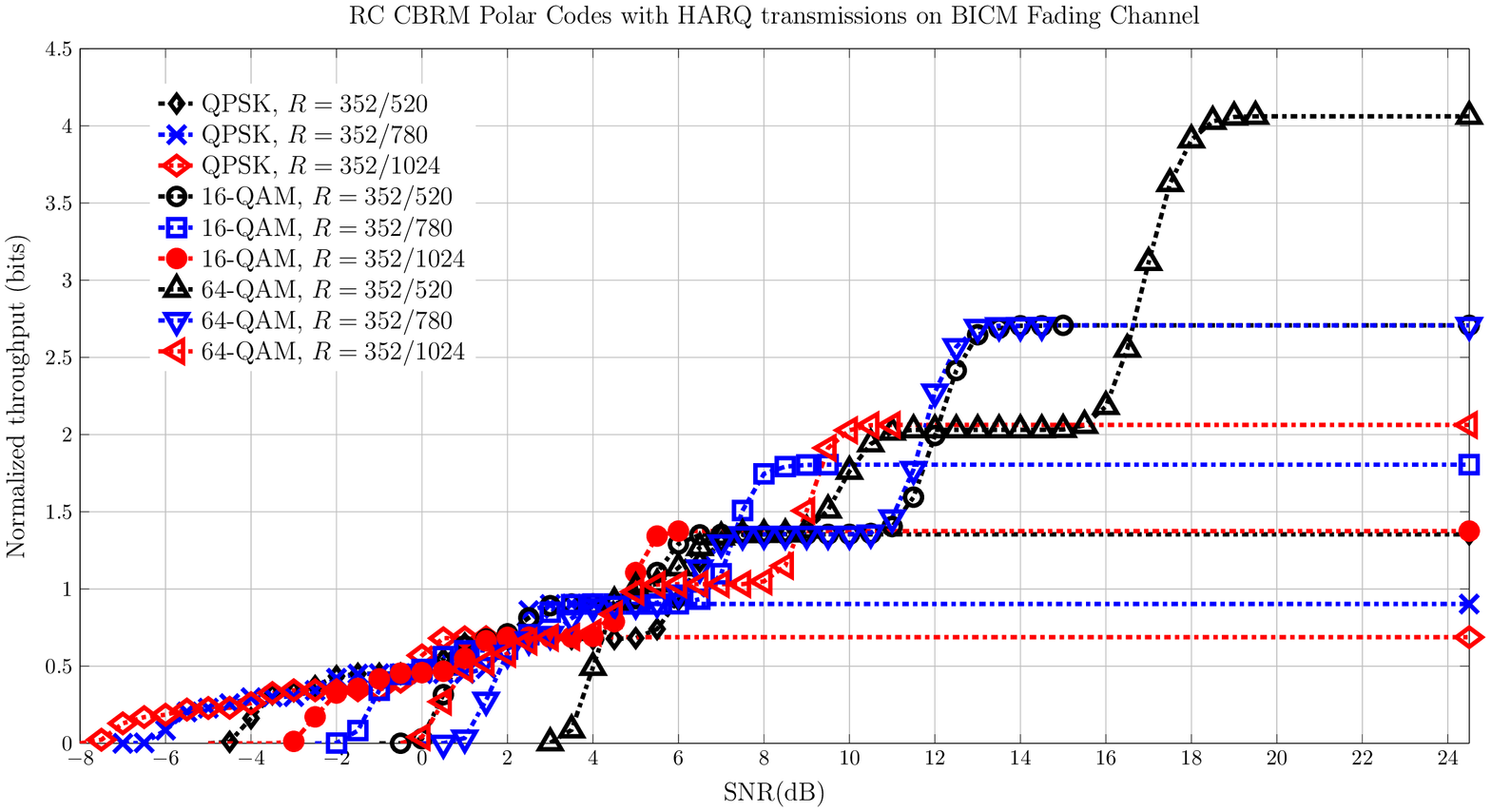}
\end{center}
\vspace{-3cm}
\caption{Adaptive modulation and coding with proposed IR HARQ RC polar codes, using QPSK, 16-QAM, or 64-QAM, on fading channels, $N=2^{12}, N'=2^{10}$, $L=32$.}
\label{fig_HARQ}
\end{figure}

\emph{Example 3:} With the same puncturing sequence for the base code of length $N'=1024$ used for comparison with turbo codes at $N=1024$, we construct a RC CBRM family of systematic polar codes with mother code block length $N=2^{14}$ and information block length  $k=4951$ bits. We show the performance at code rates $R\in \{ 0.302, 0.452, 0.6,0.72 \}$ under the circular buffer rate matching scheme corresponding to a two stage encoding $B_{2^{14}}(F_2^{\otimes 10} \otimes F_2^{\otimes 4})$. As explained before, this is equivalent to a one stage encoding, followed by appropriate arrangement of the coded bits in the circular buffer.
The information set is selected at the highest code rate $R=0.72$ in the RC family under SNR of $4$ dB based on the bit-channel error probabilities estimated by the Gaussian approximation approach.
The BER and FER decoding performances of the RC CBRM family for BPSK transmission over AWGN channel are presented in Fig. \ref{fig_RC_SP}.
The decoder employs the list SC decoding with list size $32$ and $24$ CRC bits, i.e., the $4951$ non-frozen bits consists of $4927$ information bits and $24$ CRC bits. For comparison with other codes, we also show in Fig. \ref{fig_RC_SP} the decoding performances of a rate-compatible spatially-coupled quasi-cyclic LDPC (SC-QC LDPC) code of mother code $(19656,5922)$ where the rate matching was simply done by regular puncturing of the mother code, called $C_2$ as reported in \cite{SCQCLDPC}. One can observe that although the SC QC LDPC codes have slightly better performance at the lower rate codes, the constructed RC CBRM polar code family outperforms the rate-compatible SC-QC LDPC code of \cite{SCQCLDPC} at higher rates with more than $1$ dB gain at $R=0.72$. 
From the BER performance, we note that the constructed RC polar code family has a consistent gap to the Shannon limit across all rates, which implies their capacity achieving property and the robustness of the proposed rate-compatible scheme. \fexmpl

 Next, we analyze the throughput performance of IR HARQ on BICM channels, with the designed RC polar codes.
With HARQ, the \emph{normalized throughput} for transmissions with code rate $R$ and modulation order $M$ is defined as $\mathcal{T}=R\log_2(M)(1-\mbox{BLER})/\bar{t}$, where BLER is the average residual code block error rate after all decoding attempts in the $t$ transmissions and $\bar{t}$ is the average number of transmissions required for successful decoding of an information block at the tested SNR.

\emph{Example 4:} Using the base code of length $N'=32$ and its PPA sequence shown in \emph{Example 1}, a RC compatible family of mother code length $N=1024$ and constituting of RC polar codes with rates  $R \in \{11/12, 11/16, 11/32\}$ is constructed.
The performance of IR HARQ and CC HARQ is analyzed assuming a maximum of $t=4$ transmissions on an $16$-QAM AWGN channel, and with SC decoding. Simple Chase combining is assumed across transmissions, although the performance can be improved with minimal additional complexity by schemes such as accumulator feedback \cite{el2013soft}. As expected, IR HARQ provides about 3 dB gain over CC HARQ. We observe that IR HARQ is also better than CC HARQ at $R=11/32$, although all $N$ bits are transmitted at each transmission. However, with CC the bit-mapping is the same across transmissions, starting from the first column in the PPA-permuted circular array, but with IR the bit-to-symbol mappings change across transmissions due to the circular shifting of the bit-mapping pattern at different transmissions, where the BICM bit-mapping of the $r$th transmission is applied circularly starting at the $((r-1)2^p/4+1)$th column of the permuted matrix as described in Section \ref{Sec_interleaver}. Hence, we conclude that changing the bit-to-symbol mappings across transmissions, implicitly done with our circular buffer rate matching, can provide extra diversity and improve performance. \fexmpl

\emph{Example 5:} We design a systematic RC polar code family of mother polar code with $N=2^{12}$ and $k=1408$ bits at rates $R\in \{ 0.34, 0.45, 0.67 \}$ with a maximum of $t=4$ transmissions.
The circular buffer rate matching and bit-mapping assumes the two stages encoding structure corresponding to $B_{2^{12}}(F_2^{\otimes 10} \otimes F_2^{\otimes 2})$. The information sets are obtained by genie-aided SC decoding of the codes at $R=0.67$ for different modulation schemes.  The same puncturing pattern found by PPA on the base code of length $N'=1024$ as in \emph{Example 2} is used.
Based on Section \ref{Sec_HARQ}, at the $r$th HARQ transmission with rate $R$, $2^{12}R$ bits will be sent column-wise starting from the $((r-1)2^p/4+1)$th column of the PPA-permuted circular array.  
We show the normalized throughput curves of the above IR HARQ schemes with different modulations on fast fading channels with list-CRC decoding using $24$ CRC bits and list size $32$, in Fig. \ref{fig_HARQ}.
To design an adaptive modulation and coding (AMC) scheme, one selects the modulation and coding scheme with the highest throughput at a given SNR.
For example, at an SNR of $10$ dB, it is best to transmit using $64$-QAM modulation with code rate $R=0.34$, but if SNR improves to $8$ dB the best AMC scheme is $16$-QAM with code rate $R=0.45$. \fexmpl

\section{Conclusion}\label{conclude}

To construct families of rate-compatible polar codes, a  low-complexity algorithm that progressively selects the puncturing order on a base code of short length is devised and shown to have near-optimal performance.
A practical circular buffer rate matching scheme is devised based on a two-step polarization construction, which provides flexible selection of the transmitted bits for HARQ transmissions at any desired rate by puncturing or repetition.
Whereas previous algorithms would require storing the puncturing patterns for each designed code rate and length, the proposed scheme only needs to store the puncturing sequence of a base polar code with short length, which will be used to generate the puncturing pattern at any desired code rate and length. A bit-mapping scheme for transmissions on bit-interleaved coded modulation channels is further integrated with the circular buffer rate matching scheme, while provably preserving the code polarization in the presence of punctured bits.
Our theoretical analysis shows that the
punctured polar codes can have a capacity achieving property.
With
list decoding, the performance of proposed rate-compatible polar codes is shown to be comparable to those of the LTE turbo codes and the  spatially-coupled quasi-cyclic LDPC codes, \emph{cf.} \cite{SCQCLDPC}, and does not suffer from error floors.
The proposed scheme also naturally applies as a rate matching and bit-mapping scheme for channels that do not support HARQ retransmission, such as the 3GPP new radio (NR) control channel, where the rate of transmission is adapted according to the resource elements available and the channel quality.
 We conclude that the rate-compatible polar coding schemes of this paper make the adoption of polar codes in future wireless systems more practical.

\bibliographystyle{IEEEtran}
\bibliography{RCbib2}

\begin{thebibliography}{10}
\providecommand{\url}[1]{#1}
\csname url@samestyle\endcsname
\providecommand{\newblock}{\relax}
\providecommand{\bibinfo}[2]{#2}
\providecommand{\BIBentrySTDinterwordspacing}{\spaceskip=0pt\relax}
\providecommand{\BIBentryALTinterwordstretchfactor}{4}
\providecommand{\BIBentryALTinterwordspacing}{\spaceskip=\fontdimen2\font plus
\BIBentryALTinterwordstretchfactor\fontdimen3\font minus
  \fontdimen4\font\relax}
\providecommand{\BIBforeignlanguage}[2]{{%
\expandafter\ifx\csname l@#1\endcsname\relax
\typeout{** WARNING: IEEEtran.bst: No hyphenation pattern has been}%
\typeout{** loaded for the language `#1'. Using the pattern for}%
\typeout{** the default language instead.}%
\else
\language=\csname l@#1\endcsname
\fi
#2}}
\providecommand{\BIBdecl}{\relax}
\BIBdecl

\bibitem{3GPPNR}
3GPP, ``{TR38.802, Study on New Radio (NR) Access Technology Physical Layer
  Aspects},'' 3rd Generation Partnership Project (3GPP), Technical
  Specification Group Radio Access Network, Tech. Rep., 2016.

\bibitem{el2009design}
M.~El-Khamy, J.~Hou, and N.~Bhushan, ``Design of rate-compatible structured
  {LDPC} codes for hybrid {ARQ} applications,'' \emph{IEEE Journal on Selected
  Areas in Communications}, vol.~27, no.~6, pp. 965--973, 2009.

\bibitem{DRCLDPC}
T.~V. Nguyen, A.~Nosratinia, and D.~Divsalar, ``The design of rate-compatible
  protograph {LDPC} codes,'' \emph{IEEE Trans. Commun.}, vol.~60, no.~10, pp.
  2841--2850, October 2012.

\bibitem{RPLDC}
D.~G.~M. Mitchell, M.~Lentmaier, A.~E. Pusane, and D.~J. Costello, ``Randomly
  punctured {LDPC} codes,'' \emph{IEEE Journal on Selected Areas in
  Communications}, vol.~34, no.~2, pp. 408--421, Feb 2016.

\bibitem{SCQCLDPC}
K.~Liu, M.~El-Khamy, and J.~Lee, ``Finite-length algebraic spatially-coupled
  quasi-cyclic {LDPC} codes,'' \emph{IEEE Journal on Selected Areas in
  Communications}, vol.~34, no.~2, pp. 329--344, Feb 2016.

\bibitem{Arikan_09}
E.~Ar{\i}kan, ``Channel polarization: A method for constructing
  capacity-achieving codes for symmetric binary-input memoryless channels,''
  \emph{IEEE Trans. on Inf. Theory}, vol.~55, no.~7, pp. 3051--3073, 2009.

\bibitem{arikan2009rate}
E.~Ar{\i}kan and E.~Telatar, ``On the rate of channel polarization,''
  \emph{Proc. IEEE Int. Symp. Inf. Theory}, pp. 1493--1495, 2009.

\bibitem{stolte2002Phd}
N.~Stolte, ``Rekursive codes mit der {Plotkin}-konstruktion und ihre
  decodierung,'' Ph.D. dissertation, Technische Universitat Darmstadt, Jan.
  2002, translated to ``Recursive codes with the {Plotkin}-construction and
  their decoding''.

\bibitem{el2016binary}
M.~El-Khamy, H.~Lin, and J.~Lee, ``Binary polar codes are optimised codes for
  bitwise multistage decoding,'' \emph{Electronics Letters}, vol.~52, no.~13,
  pp. 1130--1132, 2016.

\bibitem{schnabl1995soft}
G.~Schnabl and M.~Bossert, ``Soft-decision decoding of {Reed-Muller} codes as
  generalized multiple concatenated codes,'' \emph{IEEE Trans. Inf. Theory},
  vol.~41, no.~1, pp. 304--308, Jan 1995.

\bibitem{dumer2000recursive}
I.~Dumer and K.~Shabunov, ``Recursive constructions and their maximum
  likelihood decoding,'' in \emph{Proc. 38th Allerton Conf. Commun., Control,
  and Computing}, 2000, pp. 71--80.

\bibitem{list_2011}
I.~Tal and A.~Vardy, ``List decoding of polar codes,'' in \emph{Proc. IEEE Int.
  Symp. Inf. Theory (ISIT)}, July 2011, pp. 1--5.

\bibitem{Korada_Sasoglu_Urbanke_10}
S.~B. Korada, E.~Sasoglu, and R.~Urbanke, ``Polar codes: Characterization of
  exponent, bounds, and constructions,'' \emph{IEEE Trans. Inf. Theory},
  vol.~56, no.~12, pp. 6253--6264, 2010.

\bibitem{matrix_reduction_polar}
D.-M. Shin, S.-C. Lim, and K.~Yang, ``Design of length-compatible polar codes
  based on the reduction of polarizing matrices,'' \emph{IEEE Trans. Commun.},
  vol.~61, no.~7, pp. 2593--2599, 2013.

\bibitem{first_com_polar}
A.~Eslami and H.~Pishro-Nik, ``A practical approach to polar codes,'' in
  \emph{Proc. IEEE Int. Symp. on Inf. Theory (ISIT)}, 2011, pp. 16--20.

\bibitem{harq_polar}
K.~Chen, K.~Niu, and J.~Lin, ``A hybrid {ARQ} scheme based on polar codes,''
  \emph{IEEE Communications Letters,}, vol.~17, no.~10, pp. 1996--1999, 2013.

\bibitem{punc_pattern_polar}
L.~Zhang, Z.~Zhang, X.~Wang, Q.~Yu, and Y.~Chen, ``On the puncturing patterns
  for punctured polar codes,'' in \emph{in Proc. IEEE Int. Symp. Inf. Theory
  (ISIT)}, 2014, pp. 121--125.

\bibitem{HARQ_RC_polar}
M.~El-Khamy, H.-P. Lin, J.~Lee, H.~Mahdavifar, and I.~Kang, ``{HARQ}
  rate-compatible polar codes for wireless channels,'' in \emph{Proc. IEEE
  Global Communications Conference (GLOBECOM)}, Dec 2015.

\bibitem{compound_polar}
H.~Mahdavifar, M.~El-Khamy, J.~Lee, and I.~Kang, ``Compound polar codes,'' in
  \emph{Proc. IEEE Information Theory and Applications Workshop (ITA), 2013},
  2013, pp. 1--6.

\bibitem{PolarBICM}
------, ``Polar coding for bit-interleaved coded modulation,'' \emph{IEEE
  Transactions on Vehicular Technology}, vol.~65, no.~5, pp. 3115--3127, 2016.

\bibitem{relaxed_polar_code}
M.~El-Khamy, H.~Mahdavifar, G.~Feygin, J.~Lee, and I.~Kang, ``Relaxed channel
  polarization for reduced complexity polar coding,'' in \emph{Proc. IEEE
  Wireless Communications and Networking Conference (IEEE WCNC 2015)}, Mar.
  2015, pp. 219--224.

\bibitem{mahdavifar2014performance}
H.~Mahdavifar, M.~El-Khamy, J.~Lee, and I.~Kang, ``Performance limits and
  practical decoding of interleaved {Reed-Solomon} polar concatenated codes,''
  \emph{IEEE Trans. Commun.}, vol.~62, no.~5, pp. 1406--1417, 2014.

\bibitem{kailath1967divergence}
T.~Kailath, ``The divergence and {Bhattacharyya} distance measures in signal
  selection,'' \emph{IEEE Trans. Commun. Technology}, vol.~15, no.~1, pp.
  52--60, 1967.

\bibitem{Korada_Thesis_09}
S.~B. Korada, ``Polar codes for channel and source coding,'' Ph.D.
  dissertation, {\'E}COLE POLYTECHNIQUE F{\'E}D{\'E}RALE DE LAUSANNE, 2009.

\bibitem{Tal_vardy_2013}
I.~Tal and A.~Vardy, ``How to construct polar codes,'' \emph{IEEE Trans. on
  Inf. Theory}, vol.~59, no.~10, pp. 6562--6582, Oct 2013.

\bibitem{GA_polar}
P.~Trifonov, ``Efficient design and decoding of polar codes,'' \emph{IEEE
  Transactions on Communications,}, vol.~60, no.~11, pp. 3221--3227, 2012.

\bibitem{SystematicPolar}
E.~Arikan, ``Systematic polar coding,'' \emph{IEEE Commun. Letters}, vol.~15,
  no.~8, pp. 860--862, August 2011.

\bibitem{EfficientSystematic}
H.~Vangala, Y.~Hong, and E.~Viterbo, ``Efficient algorithms for systematic
  polar encoding,'' \emph{IEEE Commun. Letters}, vol.~20, no.~1, pp. 17--20,
  Jan 2016.

\bibitem{sarkis2016flexible}
G.~Sarkis, I.~Tal, P.~Giard, A.~Vardy, C.~Thibeault, and W.~J. Gross,
  ``Flexible and low-complexity encoding and decoding of systematic polar
  codes,'' \emph{IEEE Transactions on Communications}, vol.~64, no.~7, pp.
  2732--2745, 2016.

\bibitem{el2013soft}
M.~El-Khamy, J.~Lee, and I.~Kang, ``Soft turbo {HARQ} combining,'' in
  \emph{Proc. IEEE International Conference on Communications (ICC)}, 2013, pp.
  5542--5547.

\end{thebibliography}

\end{document}